\documentclass[english]{IEEEtran}
\usepackage[margin=2.2cm, centering]{geometry}
\usepackage{multirow}
\usepackage{booktabs}
\usepackage{lipsum}
\usepackage{amsfonts}
\usepackage{amsmath,cases,amssymb}
\usepackage{graphicx}
\usepackage{cite}
\usepackage{subfigure}
\usepackage{epsfig}
\usepackage{url}
\usepackage{algorithm}
\usepackage{algorithmic}
\usepackage{epstopdf}
\usepackage{balance}
\usepackage{color}
\newlength\figureheight 
\newlength\figurewidth 

\makeatletter
\floatstyle{ruled}
\newfloat{algorithm}{tbp}{loa}
\providecommand{\algorithmname}{Algorithm}
\floatname{algorithm}{\protect\algorithmname}

\usepackage{amsfonts}
\usepackage{cite}
\usepackage{array}
\usepackage{algorithm}
\usepackage{algorithmic}
\usepackage{subfigure}

\DeclareMathOperator{\tr}{tr}

\DeclareMathOperator*{\st}{subject\ to}

\newtheorem{proposition}{Proposition}

\makeatother
\makeatother
\providecommand{\remarkname}{Remark}
\providecommand{\theoremname}{Theorem}

\newcommand{\bw}{\mathbf{w}}
\newcommand{\bh}{\mathbf{h}}

\newcommand{\Pro}{\mathsf{Pr}}

\let\mybibitem\bibitem
\renewcommand{\bibitem}[1]{%
 \ifthenelse{\equal{#1}{Vartiainen2010P} \OR \equal{#1}{hoyhtya2010} \OR \equal{#1}{CHLui14} \OR \equal{#1}{GDing15} \OR \equal{#1}{LiuTMC16}}
 {\color{black}\mybibitem{#1}}
 {\color{black}\mybibitem{#1}}%
}

\begin{document}

\title{{Cooperative Prediction-and-Sensing Based Spectrum Sharing in Cognitive Radio Networks}}
\author{\IEEEauthorblockN{Van-Dinh~Nguyen and Oh-Soon~Shin}
\thanks{The authors are with the School of Electronic Engineering \& Department of  ICMC Convergence Technology, Soongsil University, Seoul 06978, Korea (e-mail: \{nguyenvandinh, osshin\}@ssu.ac.kr).}}

\maketitle
\begin{abstract}
This paper proposes prediction-and-sensing-based spectrum sharing, a new spectrum-sharing model for cognitive radio networks, with a time structure for each resource block divided into a spectrum prediction-and-sensing phase and a data transmission phase. Cooperative spectrum prediction is incorporated as a sub-phase of spectrum sensing in the first phase. We investigate a joint design of transmit beamforming at the secondary base station (BS) and sensing time. The primary design goal is to maximize the sum rate of all secondary users (SUs) subject to the minimum rate requirement for all SUs, the transmit power constraint at the secondary BS, and the interference power constraints at all primary users. The original problem is difficult to solve since it is highly nonconvex. We first convert the problem into a more tractable form, then arrive at a convex program based on an inner approximation framework, and finally propose a new algorithm to successively solve this convex program. We prove that the proposed algorithm iteratively improves the objective while guaranteeing convergence at least to local optima. Simulation results demonstrate that the proposed algorithm reaches a stationary point after only a few iterations with a substantial performance improvement over existing approaches.
\end{abstract}

\begin{IEEEkeywords} Cognitive radio, nonconvex programming, sum rate, transmit beamforming, opportunistic spectrum access, prediction accuracy, spectrum sensing, spectrum sharing,  spectrum underlay. \end{IEEEkeywords}

\section{Introduction} \label{introduction}

The ever-growing demand for mobile traffic requires new technologies to increase the data rate and enhance connectivity  using finite radio resources  \cite{Ericson13}. However, conventional static spectrum allocation policies can no longer provide  substantial improvements since these are subjected to an inefficient use of the wireless spectrum  \cite{HaykinJSAC05}. Nonetheless, the Federal Communications Commission (FCC)  \cite{FCC}  reported that the majority of primary users (PUs,  licensed users)   under-utilize  their allocated resources at  any given time and location. Therefore, cognitive radio (CR) networks, also known as dynamic spectrum access (DSA) networks \cite{SafaviTCCN15,SafaviDySPAN15}, have been proposed as  a powerful means to better utilize spectrum resources over conventional static spectrum allocation policies \cite{TragosCST13}.

In general, CR models can be further divided into two categories: opportunistic spectrum access  and spectrum underlay \cite{ZhaoSPM07}. In the former case,  secondary users (SUs, unlicensed users)  access the frequency bands only when the PUs are not transmitting \cite{GanTVT12,PehTVT09,ZhaoSPM07,ZhaoWC07,LiangTWC08,LiuTMC16}, and  interference constraints are not imposed on  SUs' transmission. Instead, the SUs need to detect the licensed frequency bands to avoid interfering with the PUs. In the latter case,   the SUs use the frequency bands even when the PUs are transmitting. However, they do so with restricted access and need to avoid causing detrimental interference to the PUs \cite{Zhang08,ZhangTSP08,NguyenTVT16,NguyenCL16,NguyenCL17,HeTWC14,Zhang09}. In addition,  sensing-based spectrum sharing was proposed in \cite{KangTVT09}, using a hybrid model of both opportunistic spectrum access  and spectrum underlay to exploit the spectrum resource more efficiently.

\subsection{Related Works}

In CR networks, spectrum sensing is a basic function and core step to enable  secondary systems to detect the spectrum holes and  states of the PUs \cite{LiangTWC08}. However, the results of the spectrum sensing are unstable and unreliable  due to the effects of multipath fading and path loss. Recently,  spectrum prediction,  which is based on the means of the historical spectrum sensing statistic, was proposed to combat the bottlenecks of spectrum sensing \cite{TumuluruWCMC12}. Two widely used prediction methods among such are hidden Markov models and neural networks 
(e.g., \noindent\cite{XingWC13} and the references therein). Yang \textit{et al.} \cite{YangCL15} proposed a redesigned frame structure-incorporated spectrum prediction to select  channels for sensing only from channels predicted to be inactive. {\color{black} Besides, long term information based on prioritization of channels  was proposed in \cite{Vartiainen2010P} to guide sensing, which helps save  computational resources, and an extension using both long- and short-term
history information was also  considered in \cite{hoyhtya2010}.  In \cite{CHLui14}, the authors proposed traffic classification
algorithms to estimate the PU traffic periods and PU traffic parameters.  The fundamental limits of predictability in radio spectrum state was studied in \cite{GDing15} to obtain spectrum awareness, which is the prerequisite to allow a SU to opportunistically access  the licensed frequency bands}. In fact, an accurate result for the spectrum prediction  is impossible to obtain due to the time-varying nature of the spectrum environment. Thus, a new spectrum prediction protocol to improve the prediction accuracy remains as an open problem.

Multi-antenna transmissions already play a key role in current-generation wireless communications and will be even more important to 5G systems and beyond.   
Transmit beamforming  improves the capacity and extends coverage for wireless communication systems without the need for additional bandwidth and/or transmit power. For transmit beamforming, the secondary base station (BS) requires  knowledge of the channels to the SUs and the PUs, which can be obtained via channel estimation. In practice, the secondary BS cannot expect to have perfect channel knowledge due to errors in the estimation or other
factors, such as quantization, thus requiring a robust beamforming design in the presence of channel uncertainty \cite{MaCL13,HuangTSP2012}. In addition, the perfect channel state information (CSI) of the PUs' channels is even more difficult to obtain at the secondary BS since two systems operate independently \cite{NguyenCL16}. The works in \cite{MaCL13} and \cite{HuangTSP2012} applied semi-definite programming (SDP) method to find the optimal solutions for complex matrices,  where rank-one constraints are omitted. Then,  a randomized approximate solution is employed to recover rank-one solutions. However, as noted in \cite{PhanTSP12}, such randomization techniques are  inefficient.

The spectral efficiency maximization, also known as  sum rate maximization,  has been a classical problem in CR networks and has been extensively studied recently \cite{Zhang08,HeTWC14,MaCL13,Zhang09,ZhangTSP08,NguyenTVT16,NguyenCL16,NguyenCL17}. Depending on the power usage, the spectral efficiency problem has been studied  with a sum power constraint  \cite{Zhang09,ZhangTSP08,NguyenCL17} and  per-antenna power constraints  \cite{NguyenTVT16,NguyenCL16}.
However, the minimum rate requirements for all SUs were not addressed in \cite{Zhang08,HeTWC14,MaCL13,Zhang09,ZhangTSP08,NguyenTVT16,NguyenCL16}, although such rate constraints are crucial to resolving the so-called user fairness. Without the minimum rate requirement for each SU, the secondary BS will favor  SUs with good channel gains by allocating a large amount of power to them. Consequently, the spectral efficiency of CR networks is mostly contributed by  SUs with good channel gains, and thus, the remaining SUs may achieve a very low throughput. Moreover, finding  a feasible point of involved optimization variables to meet the throughput constraints is also  difficult  since the feasible set is nonconvex and nonsmooth.

\subsection{Motivation and Contributions}

In this paper, we propose  prediction-and-sensing-based spectrum sharing (PSBSS), a new spectrum-sharing model for a secondary system consisting of a multi-antenna BS transmitting data to multiple SUs in the presence of multiple PUs. {\color{black}The two systems operate in the same frequency band to exploit the available spectrum more effectively}.  This is different from both opportunistic spectrum access and spectrum underlay in that if the PUs are detected to be idle, the secondary system will transmit power as long as the performance improves without any restricted power at the secondary BS, and vice versa when  the PUs are detected to be active. To improve the prediction accuracy, we redesign the time structure for each resource block with cooperative spectrum prediction between the secondary BS and all SUs. We restrict ourselves to linear beamforming strategies and consider the sum rate maximization problem subject to  the minimum rate requirements for each SU, transmit power constraints at the secondary BS, and interference power constraints at all PUs. In addition,  the channel vectors of all SUs and PUs are imperfectly known at the secondary BS, where  the CSI errors are norm-bounded.

In fact,  the optimization  problem under consideration is  highly nonconvex, and thus, the optimal solutions are  computationally difficult to find. Nevertheless, we propose a new iterative algorithm  to directly handle such a highly nonconvex problem that does not follow the SDP method due to the inefficiency  mentioned above. We also discuss its practical implementation to ensure that the proposed algorithm can be successfully solved in the first iteration. Our main contributions are summarized as follows:
\begin{itemize}
   \item We propose  cooperative spectrum prediction between all SUs and the secondary BS, which helps reduce the detection error and improve the detection accuracy.
	
\item We propose a new iterative low-complexity algorithm to obtain the computational solution of the optimization problem.  Here  the proposed design is based on an inner approximation algorithm, and we completely avoid rank-one constraints, which is different from covariance matrices \cite{MaCL13,HuangTSP2012}. Thus, the proposed algorithm requires the minimum number of optimization variables and has a moderate dimension.
  		
\item  The obtained solutions are guaranteed to locate the Karush-Kuhn-Tucker (KKT) solution to the original nonconvex program. Numerical results are also provided to demonstrate the effectiveness of the proposed algorithm, showing quite fast computation with converging in a few iterations. These results show that   the system performance of the proposed PSBSS outperforms both opportunistic spectrum access and spectrum underlay.

\end{itemize}

\subsection{Paper Organization and Notation}

The rest of the paper is organized as follows. System model is described in Section \ref{Systemmodel}. In Section \ref{PSAOPF}, we present the prediction-and-sensing analysis and optimization problem formulation. We devise the optimal solution to the sum rate maximization problem  in Section \ref{Proposedalgorithm}. Numerical results are provided in Section \ref{Numericalresults}, and Section \ref{Conclusion} concludes the paper.

\emph{Notation}: $\mathbf{x}^{H}$, $\mathbf{x}^{T}$, and $\tr(\mathbf{x})$  are the Hermitian transpose, normal transpose,   and trace of a vector $\mathbf{x}$, respectively. $\|\cdot\|$ and $|\cdot|$ denote the Euclidean norm of  vector and the absolute value of a complex scalar, respectively.  $\mathbf{x}\sim\mathcal{CN}(\boldsymbol{\eta},\boldsymbol{Z})$ means that $\mathbf{x}$ is a random vector following a complex circularly-symmetric Gaussian distribution with mean vector  $\boldsymbol{\eta}$ and covariance matrix $\boldsymbol{Z}$. $\mathbb{E}[\cdot]$ denotes the statistical expectation.   $\Re\{\cdot\}$ represents the real part of the argument. The inner product $\left\langle \mathbf{x}, \mathbf{y}\right\rangle$ is defined as $\mathrm{trace}(\mathbf{x}^H\mathbf{y})$. $\nabla$ denotes the first-order differential operator.

\section{System Model}\label{Systemmodel}
\subsection{Signal Model}

\begin{figure}[t]
\centering
\includegraphics[width=1\columnwidth,trim={-0cm -0.0cm 0cm 0cm}]{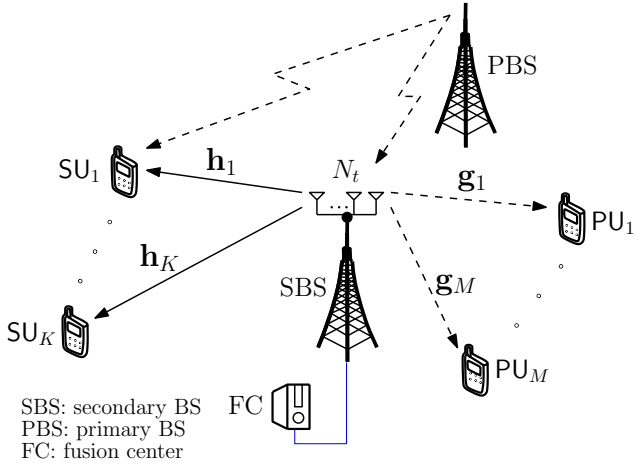}
\caption{ A CR network model with multiple SUs and
PUs. }
\label{fig:SM:1}
\end{figure}

We consider a cognitive transmission scenario where a secondary BS equipped with $N_t$ transmit antennas  serves $K$ single-antenna SUs  in the presence of $M$ single-antenna PUs,  as shown Fig.~\ref{fig:SM:1}.  We assume that two systems  operate in the same frequency band. 
The channel vectors from the secondary BS to the $k$-th SU and the $m$-th PU are represented by $\mathbf{h}_{k}\in\mathbb{C}^{N_t\times 1}$ and $\mathbf{g}_{m}\in\mathbb{C}^{N_t\times 1}$, respectively, which include the effects of the large-scale path loss and small-scale fading. We assume that $\mathbf{h}_{k}\in\mathbb{C}^{N_t\times 1},\, k\in\mathcal{K}\triangleq\{1,2,\cdots, K\}$ and $\mathbf{g}_{m}\in\mathbb{C}^{N_t\times 1},\, m\in\mathcal{M}\triangleq\{1,2,\cdots, M\}$ remain unchanged during a
transmission block and change independently from one block to another.

In the secondary system, linear beamforming  is employed at the secondary BS to transmit information signals to the SUs.
 Specifically, the information intended for the $k$-th SU, denoted by $x_k\in\mathbb{C}$ with $\mathbb{E}\{|x_k|^2\} = 1$, is multiplied by the beamformer $\mathbf{w}_{k,i}\in\mathbb{C}^{N_t\times 1}, i = \{0,1\}$. If the PUs are detected to be active $(i=1)$, the secondary BS transmits with beamformer $\mathbf{w}_{k,1}, \forall k$. If the PUs are detected to be absent $(i=0)$, the secondary BS transmits with beamformer $\mathbf{w}_{k,0}, \forall k$. Then, the received signal at the $k$-th SU depending on the PUs' channel states is given as
\begin{IEEEeqnarray}{rCl}
y_{k}^{(i)}=\bh^H_{k}\bw_{k,i}x_{k}+\sum_{j\in\mathcal{K}\backslash \{k\}}\bh^{H}_{k}\bw_{j,i}x_{j}+\beta_p^{(i)}\mathbb{I}_p+n_{k},\nonumber\\
 i = \{0,1\}\label{eq:signalmodel}\quad
\end{IEEEeqnarray}
where $\beta_p^{(1)} = 1$ if the PUs are active and $\beta_p^{(0)} = 0$ if the PUs are absent. $\mathbb{I}_p$ is referred to as the summed interference caused by the primary BS, which is assumed to be equal at all $K$ SUs. Without loss of generality, we assume the averaged interference received by each SU is $\mathbb{E}\{|\mathbb{I}_p|^2\} = \bar{\mathbb{I}}_p$ since the transmit strategies of the two systems are independent. $n_k\sim\mathcal{CN}(0,\sigma_k^2)$ is the zero-mean circularly symmetric complex Gaussian noise with variance $\sigma_k^2$. For simplicity, let us define $\bw_{0}\triangleq[\bw_{1,0}^T,\bw_{2,0}^T,\cdots,\bw_{K,0}^T]^T$ and $\bw_{1}\triangleq[\bw_{1,1}^T,\bw_{2,1}^T,\cdots,\bw_{K,1}^T]^T$.

\textit{Channel state information:} In the previous works \cite{Zhang08,ZhangTSP08,NguyenTVT16,NguyenCL16,HeTWC14,Zhang09},  the CSIs of the channel vectors $\mathbf{h}_{k}, \forall k$ and $\mathbf{g}_{m}, \forall m$  are assumed to be perfectly known at the secondary BS. To ensure a practical consideration, we assume  that the channel vectors $\mathbf{h}_{k}$ and $\mathbf{g}_{m}$ are  imperfectly known at the secondary BS as \cite{HuangTSP12,YooIT06}
\begin{IEEEeqnarray}{rCl}
f\bigl(\mathbf{h}_k\mathbf{h}_k^H - \tilde{\mathbf{h}}_k\tilde{\mathbf{h}}_k^H\bigl) &\leq& \delta_k, \forall k, \label{eq:csih}\\
f\bigl(\mathbf{g}_m\mathbf{g}_m^H - \tilde{\mathbf{g}}_m\tilde{\mathbf{g}}_m^H\bigl) &\leq& \hat{\delta}_m, \forall m \label{eq:csig}
\end{IEEEeqnarray}
where $\tilde{\mathbf{h}}_k$ and $\tilde{\mathbf{g}}_m$ are the channel estimates for the $k$-th SU and the $m$-th PU  available
at the secondary BS, respectively. $f(\boldsymbol{X}) $ is the so-called  spectral radius of matrix $\boldsymbol{X},$ i.e., $f(\boldsymbol{X}) =  \max_i|\lambda_i(\boldsymbol{X})|$ with its eigenvalues $\lambda_i(\boldsymbol{X})$.  $\delta_k$ and $\hat{\delta}_m$ represent the associated CSI errors, which are assumed to be deterministic and
bounded. Therefore, $\delta_k$ ($\hat{\delta}_m$, resp.) is the size of the uncertainty region of the estimated CSI for the $k$-th SU ($m$-th PU,  resp.). In addition, the channel uncertainties can be reformulated as \cite{HuangTSP12}
\begin{equation}
\left\{
							\begin{array}{ll}
									\delta_k = \epsilon_s\|\tilde{\mathbf{h}}_k\|^2,\quad \forall k,   \\
									\hat{\delta}_m = \epsilon_p\|\tilde{\mathbf{g}}_m\|^2,\quad \forall m
							\end{array}
			   \right.
\label{eq:CSIerrors}
\end{equation}
where $\epsilon_s$ and $\epsilon_p$ are the normalized uncertainty levels associated with the $k$-th SU and $m$-th PU, respectively. In fact, $\epsilon_s$ is much smaller than $\epsilon_p$ since the SUs are active users in the considered system.

\subsection{Prediction-and-Sensing Based Spectrum Sharing Model}

\begin{figure}[t]
\centering
\includegraphics[width=1\columnwidth,trim={-0cm -0.0cm 0cm 0cm}]{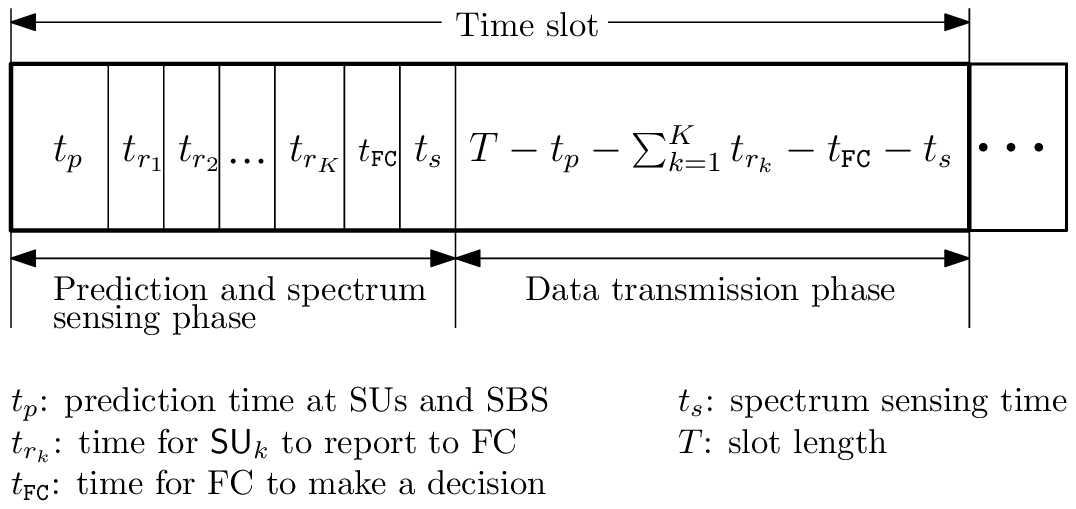}
\caption{ Time slot structure of the proposed prediction-and-sensing-based spectrum sharing in CR networks. }
\label{fig:SM:2}
\end{figure}

In our model, the secondary system first listens to the spectrum allocated to the PUs to detect their states. Then, the secondary BS decides the transmission strategies based on the detection results. Specifically, if the PUs are inactive, the SUs will transmit over the band without any restrictions in   transmit power to achieve higher system performance. If the PUs are active, the SUs   have restricted access and need to avoid causing detrimental interference to the PUs. Consequently, the secondary system can exploit the radio frequency spectrum more efficiently than when using \textit{spectrum underlay} \cite{Zhang08,ZhangTSP08,NguyenTVT16,NguyenCL16,HeTWC14,Zhang09} and \textit{opportunistic spectrum access} \cite{GanTVT12,PehTVT09,ZhaoSPM07,ZhaoWC07,LiangTWC08}.

Fig.~\ref{fig:SM:2} depicts the time slot structure of the system, consisting of a prediction and spectrum
sensing phase and a data transmission phase  in a communication time block, $T$. {\color{black}In the prediction and spectrum sensing phase, $K$ SUs and the secondary BS independently perform local spectrum prediction on the PUs' channel within the  time duration $t_p$. To ensure low computational complexity
of the end devices, the $K$ SUs report their local spectrum prediction results to the fusion center (FC) during  $K$ mini-slots  $t_{r_1}, t_{r_2}, \cdots, t_{r_{K}}$ using a dedicated control channel \cite{CichonCST16}. The FC  combines the  prediction results of both SUs and the secondary BS, and then makes a  decision regarding the PUs' channel state within the time duration $t_{\mathtt{FC}}$. With the prediction results determined previously, only the secondary BS listens to the signals sent by the PUs and performs spectrum sensing within the time duration $t_s$, which helps facilitate global resource allocation}. During the data transmission phase, as mentioned earlier, the secondary BS will transmit the data in  the remaining fraction $T - t_{pr} -t_s$ where $t_{pr} \triangleq t_p + \sum_{k=1}^{K}t_{r_k} + t_{\mathtt{FC}}$ for whatever PUs' state.

\section{Prediction-and-Sensing Analysis and Optimization Problem Formulation}\label{PSAOPF}

\subsection{Prediction-and-Sensing Analysis}

\subsubsection{Spectrum Prediction}

Inspired by the work in \cite{LiangTWC08}, the PUs' signal is modeled as a binary stochastic process, i.e., busy ($\mathcal{H}_1$) and idle ($\mathcal{H}_0$). In addition, the PUs' arrival time  is modeled as a Poisson distribution  of parameter $\lambda$, and the holding time is modeled as a binomial distribution of parameter $\mu$ \cite{MasontaST13}. Thus, the PUs' channel is predicted to be busy with a probability of $\Pro(\mathcal{H}_1) = \mu/\lambda$ and to be idle with a probability of $\Pro(\mathcal{H}_0) = 1- \mu/\lambda$.

This paper considers an imperfect spectrum prediction  \cite{TumuluruWCMC12,XingWC13,YangCL15}, i.e., the PUs' channel can be predicted to be idle when it is  actually busy.  The artificial neural network (ANN) for spectrum prediction in \cite{TumuluruWCMC12} is adopted  thanks to low energy
consumption of SUs. In particular, each SU predicts the PUs' channel state by using 
a multilayer perceptron (MLP) predictor, where the input data is the history observations and  the output is the
prediction of the future channel states.\footnote{{\color{black}The MLP predictor requires less history observations
than the  hidden Markov model (HMM) predictor to predict the future channel states. With sufficient hidden layers,  a better performance can be achieved by the MLP predictor with only one training process \cite{TumuluruWCMC12}.}} We assume that the wrong prediction probability of the true state of the PUs' channel is equivalent for all $K$ SUs and the secondary BS, denoted by $\mathcal{P}_p^w$. Similarly, $\mathcal{P}_p^s$ is assumed to be the probability of a successful prediction  of the true state of the PUs. After conducting the prediction for the local spectrum,  all $K$ SUs report their results to the FC, and then the FC combines all $(K+1)$ results from the $K$ SUs and the secondary BS to make a  decision.   We assume $k$-out-of-$(K+1)$ rule at the FC since it requires the least communication overhead \cite{ZhangICC08}. The $\mathtt{majority}$-rule is chosen as a fusion rule in this paper since it is a trade-off for two widely used fusion rules, namely the $\mathtt{OR}$-rule and $\mathtt{AND}$-rule \cite{LiangTWC08}.  {\color{black}More specifically,  the FC with the $\mathtt{OR}$-rule and $\mathtt{AND}$-rule will result in  more miss prediction of active PUs and  more loss of SUs' transmission opportunities, respectively. On the other hand, the FC with the $\mathtt{majority}$-rule can improve the probability of  successful prediction and reduce the probability of wrong prediction simultaneously. In particular, the probability of successful prediction is close to 1 and that of wrong prediction is close to 0 for a sufficiently large $K$, which is suitable for a hyper-dense small cell deployment in 5G}. Based on the $\mathtt{majority}$-rule, the FC makes  the  decision according to  the following test:
\begin{equation}
d_{\mathrm{FC}} =  \left\{
							\begin{array}{ll}
									1\, (\mathrm{busy}),\ \mathrm{if}\; \sum_{i=1}^{K+1}\alpha_i \geq \left\lceil \frac{K+1}{2} \right\rceil \\
									0\, (\mathrm{idle}),\ \mathrm{otherwise}
							\end{array}
			   \right.
\label{eq:FCtest}
\end{equation}
where $\alpha_i$ is a binary hypothesis test reported by the $K$ SUs and the secondary BS with $\alpha_i\in\{0,1\}$, and $\left\lceil \cdot \right\rceil$ denotes the ceiling function. In \eqref{eq:FCtest}, the PUs' channel is predicted to be busy if at least half of $(K+1)$ local prediction results vote the channel as occupied. Thus, the  probabilities of wrong prediction, denoted by $\mathcal{Q}_p^w$, and  successful prediction, denoted by $\mathcal{Q}_p^s$, at the FC are given by
\begin{IEEEeqnarray}{rCl}\label{eq:globalpredic}
 \mathcal{Q}_p^w &=& \sum_{i=\bigl\lceil \frac{K+1}{2} \bigr\rceil}^{ K+1 }\Bigl(\stackrel{K+1}{i}\Bigr)(\mathcal{P}_p^w)^i(1-\mathcal{P}_p^w)^{K+1 - i},  \IEEEyessubnumber \label{eq:globalpredic:false} \\
\mathcal{Q}_p^s &=& \sum_{i=\bigl\lceil \frac{K+1}{2} \bigr\rceil}^{K+1 }\Bigl(\stackrel{K+1}{i}\Bigr)(\mathcal{P}_p^s)^i(1-\mathcal{P}_p^s)^{K+1 - i}.  \IEEEyessubnumber \label{eq:globalpredic:succ} 
\end{IEEEeqnarray}
We summarize the resulting  probabilities collected for the prediction  at the FC in Table~\ref{table:prediction}.
\begin{table}[t]
\centering
\caption{The Probabilities of the True Channel State and its Prediction Result}
\label{table:prediction}
\begin{tabular}{|c|c|c|c|}
\hline
PUs' State & Prediction &  Probability &  Composite Probability\\ \hline
Idle: $\Pro(\mathcal{H}_0)$&Idle & $1-\mathcal{Q}_p^w$  & $(1-\mathcal{Q}_p^w)\Pro(\mathcal{H}_0)$ \\ 
Idle: $\Pro(\mathcal{H}_0)$&Busy & $\mathcal{Q}_p^w$ & $\mathcal{Q}_p^w\Pro(\mathcal{H}_0)$ \\ 
Busy: $\Pro(\mathcal{H}_1)$&Idle & $1-\mathcal{Q}_p^s$ & $(1-\mathcal{Q}_p^s)\Pro(\mathcal{H}_1)$ \\ 
Busy: $\Pro(\mathcal{H}_1)$&Busy & $\mathcal{Q}_p^s$ & $\mathcal{Q}_p^s\Pro(\mathcal{H}_1)$ \\ \hline
\end{tabular}
\end{table}
From Table~\ref{table:prediction}, the overall probabilities of predicting the PUs' channel to be idle $\mathcal{P}_p^0$ or busy $\mathcal{P}_p^1$ can be calculated as
\begin{IEEEeqnarray}{rCl}\label{eq:busyidle}
 \mathcal{P}_p^0 &=& (1 - \mathcal{Q}_p^w)\Pro(\mathcal{H}_0) + (1 - \mathcal{Q}_p^s)\Pro(\mathcal{H}_1), \IEEEyessubnumber \label{eq:predic:idle} \\
\mathcal{P}_p^1 &=& \mathcal{Q}_p^w\Pro(\mathcal{H}_0) + \mathcal{Q}_p^s\Pro(\mathcal{H}_1)  \IEEEyessubnumber \label{eq:predic:busy} 
\end{IEEEeqnarray}
where $\mathcal{P}_p^0$ and $\mathcal{P}_p^1$  satisfy $\mathcal{P}_p^0 = 1 - \mathcal{P}_p^1$.

\subsubsection{Prediction-and-Sensing}
After the spectrum prediction, the secondary BS  performs spectrum sensing to determine the busy/idle state of the PUs's channel based on  two hypotheses regarding whether the  PUs are active or absent \cite{LiangTWC08}. Thus,  after spectrum sensing, the following four cases can  happen:

\begin{itemize}
\item \textit{Case 1:} the PUs are absent and the sensing result is idle. The corresponding probability of the prediction-and-sensing is 
\begin{equation}\label{eq:protrue}
\mathcal{P}_{00} = \frac{(1-\mathcal{Q}_{p}^w)\Pro(\mathcal{H}_0)(1-\mathcal{P}_f)}{(1 - \mathcal{Q}_p^w)\Pro(\mathcal{H}_0) + (1 - \mathcal{Q}_p^s)\Pro(\mathcal{H}_1)}
\end{equation}
where $\mathcal{P}_f$ is referred to as the false-alarm probability  of the test statistic  by using an energy detector \cite{LiangTWC08}: 
\begin{equation}\label{eq:false-alarm}
\mathcal{P}_f = Q\left(\Bigl(\frac{\tilde{\epsilon}}{\tilde{\sigma}_n^2}-1\Bigr)\sqrt{t_s f_s}\right)
\end{equation}
with $Q(\cdot)$ being the complementary distribution function of the standard Gaussian, i.e., $Q(x) \triangleq $\linebreak    $(1/\sqrt{2\pi})\int_{x}^{\infty}\exp(-t^2/2)dt$. $\tilde{\epsilon}$, $\tilde{\sigma}_n^2$,  and $f_s$ are the detection threshold,   variance of the noise at the secondary BS,  and  the sampling frequency, respectively. Then, the number of samples is defined as $N = t_s f_s$.

\item \textit{Case 2:} the PUs are absent  but the sensing result is busy. The corresponding false-alarm probability of the prediction-and-sensing is 
\begin{equation}
\mathcal{P}_{01} = 1- \mathcal{P}_{00}.
\end{equation}

\item \textit{Case 3:} the PUs are active  but the sensing result is idle. The corresponding probability for miss-detection during prediction-and-sensing is 
\begin{equation}\label{eq:profalse}
\mathcal{P}_{10} = \frac{(1-\mathcal{Q}_{p}^s)\Pro(\mathcal{H}_1)(1-\mathcal{P}_d)}{\mathcal{Q}_p^w\Pro(\mathcal{H}_0) + \mathcal{Q}_p^s\Pro(\mathcal{H}_1)}
\end{equation}
where $\mathcal{P}_d$ is referred to as the detection probability  of the test statistic, i.e.,
\begin{equation}\label{eq:detection}
\mathcal{P}_d = Q\Bigl(\bigl(\frac{\tilde{\epsilon}}{\tilde{\sigma}_n^2}-\gamma - 1\bigr)\sqrt{t_s f_s/(2\gamma+1)}\Bigl)
\end{equation}
with $\gamma$ being the received signal-to-noise ratio (SNR) at the secondary BS.

\item \textit{Case 4:} the PUs are active and the sensing result is busy. The corresponding detection probability for prediction-and-sensing is 
\begin{equation}
\mathcal{P}_{11} = 1- \mathcal{P}_{10}.
\end{equation}
\end{itemize}

With the above results, we summarize the prediction-and-sensing results and related rate of the SUs in Table~\ref{table:Sensing}.
\begin{table}[t]
\centering
\caption{The Probabilities of the  True Channel State, Prediction-and-Sensing Result, and Related Rate of  SUs}
\label{table:Sensing}
{\setlength{\tabcolsep}{0.2em}
\setlength{\extrarowheight}{0.4em}
\begin{tabular}{|c|c|c|c|c|c|}
\hline
PUs' State & \normalsize$\substack{\text{Prediction-}\\ \text{and-Sensing}}$ & Probability & $\normalsize\substack{\text{ Composite}\\ \text{Probability}}$ & Related Rate \\ \hline
Idle: $\Pro(\mathcal{H}_0)$ &Idle & $\mathcal{P}_{00}$  & $\Pro(\mathcal{H}_0)\mathcal{P}_{00}$ & $R_{k}^{00}(\bw_{0})$\\ 
Idle: $\Pro(\mathcal{H}_0)$&Busy & $1-\mathcal{P}_{00}$ & $\Pro(\mathcal{H}_0)(1-\mathcal{P}_{00})$ & $R_{k}^{01}(\bw_{1})$\\ 
Busy: $\Pro(\mathcal{H}_1)$&Idle & $\mathcal{P}_{10}$ & $\Pro(\mathcal{H}_1)\mathcal{P}_{10}$ & $R_{k}^{10}(\bw_{0})$\\ 
Busy: $\Pro(\mathcal{H}_1)$&Busy & $1-\mathcal{P}_{10}$ & $\Pro(\mathcal{H}_1)(1-\mathcal{P}_{10})$&  $R_{k}^{11}(\bw_{1})$\\ \hline
\end{tabular}}
\end{table}
By incorporating  the channel uncertainties in \eqref{eq:signalmodel}, the worst-case information rates  in nat/sec/Hz   for the $k$-th SU listed in Table~\ref{table:Sensing} are given as
\begin{IEEEeqnarray}{rCl}\label{eq:Rate}
R_{k}^{00}(\bw_{0}) &=& \ln\biggl(1+\frac{|\tilde{\bh}^H_{k}\bw_{k,0}|^2 - \delta_k\|\bw_{k,0}\|^2}{\chi_k^{00}(\bw_{0})}\biggr), \label{eq:R00} \\
R_{k}^{01}(\bw_{1}) &=& \ln\biggl(1+\frac{|\tilde{\bh}^H_{k}\bw_{k,1}|^2 - \delta_k\|\bw_{k,1}\|^2}{\chi_{k}^{01}(\bw_{1})}\biggr),\label{eq:R01}\\
R_{k}^{10}(\bw_{0}) &=& \ln\biggl(1+\frac{|\tilde{\bh}^H_{k}\bw_{k,0}|^2 - \delta_k\|\bw_{k,0}\|^2}{\chi_{k}^{10}(\bw_{0})}\biggr), \label{eq:R10} \\
R_{k}^{11}(\bw_{1}) &=& \ln\biggl(1+\frac{|\tilde{\bh}^H_{k}\bw_{k,1}|^2 - \delta_k\|\bw_{k,1}\|^2}{\chi_{k}^{11}(\bw_{1})}\biggr)\label{eq:R11}
\end{IEEEeqnarray}
where $\chi_k^{00}(\bw_{0})$, $\chi_{k}^{01}(\bw_{1})$, $\chi_{k}^{10}(\bw_{0})$, and $\chi_{k}^{11}(\bw_{1})$ are defined as
\begin{IEEEeqnarray}{rCl}
\chi_k^{00}(\bw_{0}) &\triangleq& \sum_{j\in\mathcal{K}\backslash \{k\}}|\tilde{\bh}^{H}_{k}\bw_{j,0}|^2 + \sum_{j\in\mathcal{K}\backslash \{k\}}\delta_k\|\bw_{j,0}\|^2 + \sigma_k^2,  \nonumber\\
\chi_{k}^{01}(\bw_{1}) &\triangleq& \sum_{j\in\mathcal{K}\backslash \{k\}}|\tilde{\bh}^{H}_{k}\bw_{j,1}|^2 + \sum_{j\in\mathcal{K}\backslash \{k\}}\delta_k\|\bw_{j,1}\|^2 + \sigma_k^2,\nonumber\\
\chi_{k}^{10}(\bw_{0}) &\triangleq&\sum_{j\in\mathcal{K}\backslash \{k\}}|\tilde{\bh}^{H}_{k}\bw_{j,0}|^2 + \sum_{j\in\mathcal{K}\backslash \{k\}}\delta_k\|\bw_{j,0}\|^2 \nonumber\\ 
&&\qquad +\; \bar{\mathbb{I}}_p + \sigma_k^2, \nonumber
\end{IEEEeqnarray}
\begin{IEEEeqnarray}{rCl}
\chi_{k}^{11}(\bw_{1}) &\triangleq& \sum_{j\in\mathcal{K}\backslash \{k\}}|\tilde{\bh}^{H}_{k}\bw_{j,1}|^2 + \sum_{j\in\mathcal{K}\backslash \{k\}}\delta_k\|\bw_{j,1}\|^2 \nonumber\\
&&\qquad +\; \bar{\mathbb{I}}_p+\sigma_k^2.\nonumber
\end{IEEEeqnarray}

\subsection{Optimization Problem Formulation}
After  the prediction and spectrum sensing phase, the secondary BS will determine  the transmission strategy according to the  prediction-and-sensing results.  From Table~\ref{table:Sensing}, the effective  rate of the $k$-th SU for prediction-and-sensing based spectrum sharing is given by
\begin{IEEEeqnarray}{rCl}\label{eq:rateachSU}
&&R_k(\bw_{0},\bw_{1},t_s)\; = \left(1 - \frac{t_{pr}+t_s}{T}\right)\Bigl[ \widetilde{\mathcal{P}}_{00}R_{k}^{00}(\bw_{0})  \nonumber\\
&&\quad +\;  \widetilde{\mathcal{P}}_{01}R_{k}^{01}(\bw_{1}) + \widetilde{\mathcal{P}}_{10}R_{k}^{10}(\bw_{0}) +  \widetilde{\mathcal{P}}_{11}R_{k}^{11}(\bw_{1})   \Bigr].\qquad
\end{IEEEeqnarray}
where $\widetilde{\mathcal{P}}_{00} = \Pro(\mathcal{H}_0)\mathcal{P}_{00}$, $\widetilde{\mathcal{P}}_{01} = \Pro(\mathcal{H}_0)(1-\mathcal{P}_{00})$, $\widetilde{\mathcal{P}}_{10}=\Pro(\mathcal{H}_1)\mathcal{P}_{10}$, and $\widetilde{\mathcal{P}}_{11}=\Pro(\mathcal{H}_1)(1-\mathcal{P}_{10})$. 
Moreover, the performance measure of interest is  the  sum rate of all SUs. Thus, the objective function can be mathematically expressed as
\begin{equation}\label{eq:obj}
R(\bw_{0},\bw_{1},t_s) \triangleq \sum_{k=1}^K R_k(\bw_{0},\bw_{1},t_s).
\end{equation}
As observed in \cite{LiangTWC08}, the sensing time $t_s$ is also incorporated as an optimization variable.  In particular, for a given target  detection probability $\mathcal{P}_d = \bar{\mathcal{P}}_d$, the  false-alarm probability $\mathcal{P}_f$ is calculated as \cite{LiangTWC08}
\begin{equation}\label{eq:falsepro}
\mathcal{P}_f = 	Q\Bigl(\sqrt{2\gamma + 1}Q^{-1}(\bar{\mathcal{P}}_d) + \sqrt{t_sf_s}\gamma\Bigr),
\end{equation}
which readily shows that a higher $t_s$ leads to a lower $\mathcal{P}_f$ and a higher $\mathcal{P}_d$, and  thus improving the system performance. However, increasing $t_s$ also negatively impacts  the system performance by reducing the time fraction for the data transmission. Consequently, from \eqref{eq:false-alarm} and \eqref{eq:detection}, the following constraint is considered:
\begin{equation}\label{eq:sensigntime}
t_s \geq \frac{1}{\gamma^2f_s}\left[Q^{-1}(\mathcal{P}_f) - Q^{-1}(\mathcal{P}_d)\sqrt{2\gamma+1}\right]^2.
\end{equation}

In this paper, the aim is to maximize the  sum rate of all SUs by jointly deriving the beamforming vectors and sensing time  under  the minimum rate requirements for each SU, the transmit power constraints at the secondary BS, and the interference power constraints at the PUs. In particular, we consider the following optimization problem:
\begin{IEEEeqnarray}{rCl}\label{eq:problem:1}
&&\underset{\bw_{0},\bw_{1},t_s}{\mathrm{maximize}}\  R(\bw_{0},\bw_{1},t_s)\IEEEyessubnumber \label{eq:problem:a}\\
                         && \st \; R_k(\bw_{0},\bw_{1},t_s) \geq \mathsf{\bar{R}}_{k},\;\forall k \in\mathcal{K},\IEEEyessubnumber \label{eq:problem:b} \\
							&&\Bigl(1 - \frac{t_{pr}+t_s}{T}\Bigl)\sum_{k=1}^K\bigl( \widehat{\mathcal{P}}_0\|\bw_{k,0}\|^2 
													 + \widehat{\mathcal{P}}_1 \|\bw_{k,1}\|^2\bigl)\leq P_{sbs},\IEEEyessubnumber \label{eq:problem:d}\\
							&&  \Bigl(1 - \frac{t_{pr}+t_s}{T}\Bigl)\sum_{k=1}^K\Bigl(\mathcal{P}_{10}\bigl(\|\tilde{\mathbf{g}}^H_m\bw_{k,0}\|^2 + \hat{\delta}_m\|\bw_{k,0}\|^2\bigr) \nonumber\\
							&&+ (1-\mathcal{P}_{10})\bigl(\|\tilde{\mathbf{g}}^H_m\bw_{k,1}\|^2 + \hat{\delta}_m\|\bw_{k,1}\|^2\bigr)\Bigl)\leq \mathcal{I}_m,\,\forall m,\IEEEyessubnumber \label{eq:problem:e} \qquad\;\\
							&& t_s \geq \frac{1}{\gamma^2f_s}\left[Q^{-1}(\mathcal{P}_f) - Q^{-1}(\mathcal{P}_d)\sqrt{2\gamma+1}\right]^2.\IEEEyessubnumber \label{eq:problem:c}
\end{IEEEeqnarray}
Constraint \eqref{eq:problem:b} requires that the minimum rate achieved by the $k$-th SU be greater than the target threshold $\mathsf{\bar{R}}_{k}$. Constraint 
\eqref{eq:problem:d} caps the total transmit power of the secondary BS at a predefined value $P_{sbs}$ with $\widehat{\mathcal{P}}_0 = \widetilde{\mathcal{P}}_{00} + \widetilde{\mathcal{P}}_{10}$ and $\widehat{\mathcal{P}}_1 = \widetilde{\mathcal{P}}_{01} + \widetilde{\mathcal{P}}_{11}$.
The last constraint \eqref{eq:problem:e} imposes the  interference power caused by the secondary BS at the $m$-th PU incorporating the channel uncertainties as less than a  predefined threshold $\mathcal{I}_m$ only when the PUs are active within the time duration $1 - \frac{t_{pr}+t_s}{T}$.

\section{Proposed Optimal Solution}\label{Proposedalgorithm}

Note that problem \eqref{eq:problem:1} is highly nonconvex, with an objective function \eqref{eq:problem:a} that is non-concave and the constraints \eqref{eq:problem:b}, \eqref{eq:problem:d}, and \eqref{eq:problem:e} that are nonconvex due to coupling between the beamforming vectors $(\bw_{0},\bw_{1})$ and sensing time $t_s$.
In this section, we solve \eqref{eq:problem:1} by developing an efficient iterative algorithm based on an inner approximation framework.

Let us start by introducing a new variable $\tau$ and making the following change of variable:
\begin{equation}\label{eq:changevariable}
1 - \frac{t_{pr}+t_s}{T} = \frac{1}{\tau},
\end{equation}
with an additional linear constraint
\begin{equation}\label{eq:addconvex}
\tau > 1.
\end{equation}
We now  equivalently express \eqref{eq:problem:1} as 
\begin{IEEEeqnarray}{rCl}\label{eq:equiexpress:1}
&&\underset{\bw_{0},\bw_{1},\tau}{\mathrm{maximize}}\  \sum\nolimits_{k=1}^K \Bigl[\widetilde{\mathcal{P}}_{00}\frac{R_{k}^{00}(\bw_{0})}{\tau} +  \widetilde{\mathcal{P}}_{01}\frac{R_{k}^{01}(\bw_{1})}{\tau}\nonumber\\
  &&\qquad +\; \widetilde{\mathcal{P}}_{10}\frac{R_{k}^{10}(\bw_{0})}{\tau} + \widetilde{\mathcal{P}}_{11} \frac{R_{k}^{11}(\bw_{1})}{\tau}\Bigl]       \IEEEyessubnumber \label{eq:eq:equiexpress:a}\\
                        &&  \st \;  \widetilde{\mathcal{P}}_{00}\frac{R_{k}^{00}(\bw_{0})}{\tau} +  \widetilde{\mathcal{P}}_{01}\frac{R_{k}^{01}(\bw_{1})}{\tau} \nonumber\\
							&& +\; \widetilde{\mathcal{P}}_{10}\frac{R_{k}^{10}(\bw_{0})}{\tau} + \widetilde{\mathcal{P}}_{11} \frac{R_{k}^{11}(\bw_{1})}{\tau} \geq \mathsf{\bar{R}}_{k},\;\forall k \in\mathcal{K},\IEEEyessubnumber \label{eq:eq:equiexpress:b} \\
							&&\sum\nolimits_{k=1}^K\Bigl(\widehat{\mathcal{P}}_0\|\bw_{k,0}\|^2  +  \widehat{\mathcal{P}}_0\|\bw_{k,1}\|^2\Bigl)\leq \tau P_{sbs},\IEEEyessubnumber \label{eq:eq:equiexpress:c}\\
							&& \sum\nolimits_{k=1}^K\Bigl(\mathcal{P}_{10}\bigl(\|\tilde{\mathbf{g}}^H_m\bw_{k,0}\|^2 + \hat{\delta}_m\|\bw_{k,0}\|^2\bigr)+ (1-\mathcal{P}_{10}) \nonumber\\
							&&\times\bigl(\|\tilde{\mathbf{g}}^H_m\bw_{k,1}\|^2 + \hat{\delta}_m\|\bw_{k,1}\|^2\bigr)\Bigl)\leq \tau\mathcal{I}_m,\forall m\in\mathcal{M},\IEEEyessubnumber \label{eq:eq:equiexpress:d} \qquad\quad\ \\
							&& \tau  \geq \frac{T}{T - t_{pr} - \Omega(\mathcal{P}_f,\mathcal{P}_d,\gamma)}\IEEEyessubnumber \label{eq:eq:equiexpress:e}
\end{IEEEeqnarray}
where $\Omega(\mathcal{P}_f,\mathcal{P}_d,\gamma) \triangleq\frac{1}{\gamma^2f_s}\bigr[Q^{-1}(\mathcal{P}_f)- Q^{-1}(\mathcal{P}_d)$ $\sqrt{2\gamma+1}\bigl]^2$. Note that  \eqref{eq:eq:equiexpress:e} also admits the linear constraint \eqref{eq:addconvex}. Interestingly,  the constraints \eqref{eq:eq:equiexpress:c} and \eqref{eq:eq:equiexpress:d} become convex with these transformations. Thus, from now on, we will consider the equivalent problem \eqref{eq:equiexpress:1} instead of \eqref{eq:problem:1} in the original form. Now, we only deal with the non-concave objective \eqref{eq:eq:equiexpress:a} and the nonconvex constraint \eqref{eq:eq:equiexpress:b}.

\textit{Concave approximation of the  objective \eqref{eq:eq:equiexpress:a}:} Let us treat the  term $R_{k}^{00}(\bw_{0})/\tau$ first. Here we need to find a concave lower bound approximation of $R_{k}^{00}(\bw_{0})/\tau$ at the $n$-th iteration in the proposed algorithm presented shortly. Thus,  we develop a lower bounding concave function for $R_{k}^{00}(\bw_{0})/\tau$. To start with, \eqref{eq:R00} can be equivalently replaced by \cite{WES06} 
\begin{IEEEeqnarray}{rCl}\label{eq:R00:1}
R_{k}^{00}(\bw_{0}) = \ln\Biggl(1+\frac{\bigl(\Re\{\tilde{\bh}^H_{k}\bw_{k,0}\}\bigr)^2 - \delta_k\|\bw_{k,0}\|^2}{\chi_k^{00}(\bw_{0})}\Biggr).\qquad
\end{IEEEeqnarray}
Let $\varphi_k^{00}(\bw_{0})\triangleq \frac{\chi_k^{00}(\bw_{0})}{\bigl(\Re\{\tilde{\bh}^H_{k}\bw_{k,0}\}\bigr)^2 - \delta_k\|\bw_{k,0}\|^2}$. Then, \eqref{eq:R00:1} becomes $R_{k}^{00}(\bw_{0}) = \ln\bigl(1 + 1/ \varphi_k^{00}(\bw_{0}) \bigr)$.  $R_{k}^{00}(\bw_{0})/\tau = \ln\bigl(1 + 1/ \varphi_k^{00}(\bw_{0}) \bigr)/\tau$ is convex in the domain $(\varphi_k^{00}(\bw_{0}) > 0, \tau >1)$ \cite{Stephen}, which can be verified by examining its Hessian. Consequently, it is useful to
develop an inner approximation of $R_{k}^{00}(\bw_{0})/\tau$. Specifically, at the feasible point $(\bw_{0}^{(n)},\tau^{(n)})$, a global lower bound of $R_{k}^{00}(\bw_{0})/\tau$ can be found as \cite{Tuybook}
\begin{IEEEeqnarray}{rCl}\label{eq:R00:2}
&&\frac{\ln\Bigl(1 + \frac{1}{\varphi_k^{00}(\bw_{0})} \Bigr)}{\tau}\geq \frac{R_{k}^{00}\bigl(\bw_{0}^{(n)}\bigr)}{\tau^{(n)}} + \Biggl\langle \nabla \frac{\ln\Bigl(1 + \frac{1}{\varphi_k^{00}(\bw_{0}^{(n)})} \Bigr)}{\tau^{(n)}},
 \nonumber\\
&&\qquad\qquad\qquad\qquad\quad\bigl(\varphi_k^{00}(\bw_{0}),\tau\bigl) - \bigl(\varphi_k^{00}(\bw_{0}^{(n)}),\tau^{(n)}\bigl)  \Biggr\rangle
\nonumber\\
&&= \mathcal{A}\bigl(\varphi_k^{00}(\bw_{0}^{(n)})\bigr)    - \mathcal{B}\bigl(\varphi_k^{00}(\bw_{0}^{(n)})\bigr) \varphi_k^{00}(\bw_{0}) \nonumber\\
&&\qquad\qquad\qquad\quad\ -\; \mathcal{C}\bigl(\varphi_k^{00}(\bw_{0}^{(n)})\bigr)\tau
\end{IEEEeqnarray}
where $\mathcal{A}\bigl(\varphi_k^{00}(\bw_{0}^{(n)})\bigr)$, $\mathcal{B}\bigl(\varphi_k^{00}(\bw_{0}^{(n)})\bigr)$, and $\mathcal{C}\bigl(\varphi_k^{00}(\bw_{0}^{(n)})\bigr)$ are defined as
\begin{IEEEeqnarray}{rCl}\label{eq:R00:3}
\mathcal{A}\bigl(\varphi_k^{00}(\bw_{0}^{(n)})\bigr) &\triangleq& 2\frac{\ln\bigl(1 + 1/ \varphi_k^{00}(\bw_{0}^{(n)}) \bigr)}{\tau^{(n)}} \nonumber\\
&&\qquad +\; \frac{1}{\tau^{(n)}\bigl(1+\varphi_k^{00}(\bw_{0}^{(n)})\bigr)} > 0,\nonumber\\
\mathcal{B}\bigl(\varphi_k^{00}(\bw_{0}^{(n)})\bigr) &\triangleq& \frac{1}{\tau^{(n)}\varphi_k^{00}(\bw_{0}^{(n)})\bigl(1+\varphi_k^{00}(\bw_{0}^{(n)})\bigr)} > 0,\nonumber\\
\mathcal{C}\bigl(\varphi_k^{00}(\bw_{0}^{(n)})\bigr) &\triangleq& \frac{\ln\bigl(1 + 1/ \varphi_k^{00}(\bw_{0}^{(n)}) \bigr)}{\bigl(\tau^{(n)}\bigr)^2}  > 0.
\end{IEEEeqnarray}
Due to the convexity of $\bigr(\Re\{\tilde{\bh}^H_{k}\bw_{k,0}\}\bigr)^2$, the first-order approximation of $\bigr(\Re\{\tilde{\bh}^H_{k}\bw_{k,0}\}\bigr)^2$ at a feasible point $\bw_{k,0}^{(n)}$ is $2\Re\{\tilde{\bh}^H_{k}\bw_{k,0}^{(n)}\}\Re\{\tilde{\bh}^H_{k}\bw_{k,0}\} - \bigr(\Re\{\tilde{\bh}^H_{k}\bw_{k,0}^{(n)}\}\bigr)^2$. Then, \eqref{eq:R00:2} can be re-expressed as
\begin{IEEEeqnarray}{rCl}\label{eq:R00:4}
&&\frac{\ln\bigl(1 + 1/ \varphi_k^{00}(\bw_{0}) \bigr)}{\tau}\geq \mathcal{A}\bigl(\varphi_k^{00}(\bw_{0}^{(n)})\bigr)  - \mathcal{B}\bigl(\varphi_k^{00}(\bw_{0}^{(n)})\bigr)\times\nonumber\\
&& \frac{\chi_k^{00}(\bw_{0})}{2\Re\{\tilde{\bh}^H_{k}\bw_{k,0}^{(n)}\}\Re\{\tilde{\bh}^H_{k}\bw_{k,0}\} - \bigr(\Re\{\tilde{\bh}^H_{k}\bw_{k,0}^{(n)}\}\bigr)^2-\delta_k\|\bw_{k,0}\|^2}\nonumber\\
&&\quad -\; \mathcal{C}\bigl(\varphi_k^{00}(\bw_{0}^{(n)})\bigr) \tau\nonumber\\
&&:=\mathcal{R}_{k}^{00,(n)}(\bw_{0},\tau)
\end{IEEEeqnarray}
over  the trust regions
\begin{IEEEeqnarray}{rCl}\label{eq:trustregion}
&&2\Re\{\tilde{\bh}^H_{k}\bw_{k,0}\} - \Re\{\tilde{\bh}^H_{k}\bw_{k,0}^{(n)}\} > 0, \forall k\in\mathcal{K},  \IEEEyessubnumber\\
&&2\Re\{\tilde{\bh}^H_{k}\bw_{k,0}^{(n)}\}\Re\{\tilde{\bh}^H_{k}\bw_{k,0}\} - \bigr(\Re\{\tilde{\bh}^H_{k}\bw_{k,0}^{(n)}\}\bigr)^2 \nonumber\\
&&\qquad\qquad\qquad\quad  -\; \delta_k\|\bw_{k,0}\|^2 >0, \forall k\in\mathcal{K}.\IEEEyessubnumber
\end{IEEEeqnarray}
Note that the inequality in \eqref{eq:R00:4} becomes the equality at optimum, i.e., 
\begin{IEEEeqnarray}{rCl}\label{eq:R00:5}
\frac{\ln\Bigl(1 + 1/ \varphi_k^{00}\bigl(\bw_{0}^{(n)}\bigl) \Bigr)}{\tau^{(n)}} = \mathcal{R}_{k}^{00,(n)}\bigr(\bw_{0}^{(n)},\tau^{(n)}\bigl).
\end{IEEEeqnarray}
In order to solve $\mathcal{R}_{k}^{00,(n)}(\bw_{0},\tau)$ by existing solvers, we further transform \eqref{eq:R00:4} to the following concave function
\begin{IEEEeqnarray}{rCl}\label{eq:R00:6}
\frac{\ln\Bigl(1 + \frac{1}{\varphi_k^{00}(\bw_{0})} \Bigr)}{\tau}&\geq& \mathcal{A}\bigl(\varphi_k^{00}(\bw_{0}^{(n)})\bigr) - \mathcal{B}\bigl(\varphi_k^{00}(\bw_{0}^{(n)})\bigr) \vartheta_k^{00} \nonumber\\
&&\quad 
-\; \mathcal{C}\bigl(\varphi_k^{00}(\bw_{0}^{(n)})\bigr) \tau\nonumber\\
&:=&\widetilde{\mathcal{R}}_{k}^{00,(n)}(\tau, \vartheta_k^{00} )
\end{IEEEeqnarray}
with additional convex quadratic constraints
\begin{IEEEeqnarray}{rCl}\label{eq:R00:7}
\delta_k\|\bw_{k,0}\|^2 \leq2\Re\{\tilde{\bh}^H_{k}\bw_{k,0}^{(n)}\}\Re\{\tilde{\bh}^H_{k}\bw_{k,0}\} \nonumber\\
\quad -\; \bigr(\Re\{\tilde{\bh}^H_{k}\bw_{k,0}^{(n)}\}\bigr)^2- \omega_k^{0}, \; \forall k,\IEEEyessubnumber \label{eq:R00:7a}\\
\Bigl(\sum_{j\in\mathcal{K}\backslash \{k\}}|\tilde{\bh}^{H}_{k}\bw_{j,0}|^2 + \sum_{j\in\mathcal{K}\backslash \{k\}}\delta_k\|\bw_{j,0}\|^2 \nonumber\\ \qquad +\; \sigma_k^2 \Bigr)/\omega_k^{0}\leq  \vartheta_k^{00}, \; \forall k\IEEEyessubnumber \label{eq:R00:7b}
\end{IEEEeqnarray}
where $\omega_k^{0}$ and $ \vartheta_k^{00}$ are newly introduced variables. The equivalence between \eqref{eq:R00:4} and \eqref{eq:R00:6} can be readily verified from the fact that the constraints \eqref{eq:R00:7a} and \eqref{eq:R00:7b} hold with equality at optimum. Thus, we can iteratively replace $R_{k}^{00}(\bw_{0})/\tau$ by $\widetilde{\mathcal{R}}_{k}^{00,(n)}(\tau, \vartheta_k^{00} )$ to achieve a concave approximation at the $n$-th iteration.

Let us define the following functions:
\begin{IEEEeqnarray}{rCl}\label{eq:SINRchanged}
\varphi_k^{01}(\bw_{1})&\triangleq& \frac{\chi_k^{01}(\bw_{1})}{|\tilde{\bh}^H_{k}\bw_{k,1}|^2 - \delta_k\|\bw_{k,1}\|^2},\nonumber \\
\varphi_k^{10}(\bw_{0})&\triangleq& \frac{\chi_k^{10}(\bw_{0})}{|\tilde{\bh}^H_{k}\bw_{k,0}|^2 - \delta_k\|\bw_{k,0}\|^2},\nonumber \\
\varphi_k^{11}(\bw_{1})&\triangleq& \frac{\chi_k^{11}(\bw_{1})}{|\tilde{\bh}^H_{k}\bw_{k,1}|^2 - \delta_k\|\bw_{k,1}\|^2}.\nonumber 
\end{IEEEeqnarray}
By following similar steps from \eqref{eq:R00:1} to \eqref{eq:R00:7}, the non-concave functions $R_{k}^{01}(\bw_{1})/\tau$, $R_{k}^{10}(\bw_{0})/\tau$, and $R_{k}^{11}(\bw_{1})/\tau$ can be iteratively replaced, respectively, by 
\begin{IEEEeqnarray}{rCl}\label{eq:Rateobj}
\widetilde{\mathcal{R}}_{k}^{01,(n)}(\tau, \vartheta_k^{01} )&:=& \mathcal{A}\bigl(\varphi_k^{01}(\bw_{1}^{(n)})\bigr)  - \mathcal{B}\bigl(\varphi_k^{01}(\bw_{1}^{(n)})\bigr) \vartheta_k^{01} \nonumber \quad\\ &&-\; \mathcal{C}\bigl(\varphi_k^{01}(\bw_{1}^{(n)})\bigr) \tau,  \label{eq:R01:concave}\\
\widetilde{\mathcal{R}}_{k}^{10,(n)}(\tau, \vartheta_k^{10} )&:=& \mathcal{A}\bigl(\varphi_k^{10}(\bw_{0}^{(n)})\bigr)  - \mathcal{B}\bigl(\varphi_k^{10}(\bw_{0}^{(n)})\bigr) \vartheta_k^{10} \nonumber \\&& -\; \mathcal{C}\bigl(\varphi_k^{10}(\bw_{0}^{(n)})\bigr) \tau,  \label{eq:R10:concave}\\
\widetilde{\mathcal{R}}_{k}^{11,(n)}(\tau, \vartheta_k^{11} )&:=& \mathcal{A}\bigl(\varphi_k^{11}(\bw_{1}^{(n)})\bigr)  - \mathcal{B}\bigl(\varphi_k^{11}(\bw_{1}^{(n)})\bigr) \vartheta_k^{11} \nonumber \\&& -\; \mathcal{C}\bigl(\varphi_k^{11}(\bw_{1}^{(n)})\bigr) \tau  \label{eq:R11:concave}
\end{IEEEeqnarray}
over  the trust regions
\begin{IEEEeqnarray}{rCl}\label{eq:trustregions}
&&2\Re\{\tilde{\bh}^H_{k}\bw_{k,1}\} - \Re\{\tilde{\bh}^H_{k}\bw_{k,1}^{(n)}\} > 0, \forall k\in\mathcal{K},\IEEEyessubnumber \\
&&2\Re\{\tilde{\bh}^H_{k}\bw_{k,1}^{(n)}\}\Re\{\tilde{\bh}^H_{k}\bw_{k,1}\} - (\Re\{\tilde{\bh}^H_{k}\bw_{k,1}^{(n)}\})^2 \nonumber \\
&&\qquad\qquad\qquad\quad  -\; \delta_k\|\bw_{k,1}\|^2 > 0, \forall k\in\mathcal{K},\IEEEyessubnumber 
\end{IEEEeqnarray}
with additional convex quadratic constraints
\begin{IEEEeqnarray}{rCl}\label{eq:Rremain}
\delta_k\|\bw_{k,1}\|^2 \leq2\Re\{\tilde{\bh}^H_{k}\bw_{k,1}^{(n)}\}\Re\{\tilde{\bh}^H_{k}\bw_{k,1}\} \nonumber\\ 
-\; \bigr(\Re\{\tilde{\bh}^H_{k}\bw_{k,1}^{(n)}\}\bigr)^2- \omega_k^{1}, \; \forall k,\IEEEyessubnumber \label{eq:Rremain:a}\\
\Bigl(\sum_{j\in\mathcal{K}\backslash \{k\}}|\tilde{\bh}^{H}_{k}\bw_{j,1}|^2 + \sum_{j\in\mathcal{K}\backslash \{k\}}\delta_k\|\bw_{j,1}\|^2  \nonumber\\
+\; \sigma_k^2 \Bigr)/\omega_k^{1}\leq  \vartheta_k^{01}, \; \forall k,\IEEEyessubnumber \label{eq:Rremain:b}\\
\Bigl(\sum_{j\in\mathcal{K}\backslash \{k\}}|\tilde{\bh}^{H}_{k}\bw_{j,0}|^2 + \sum_{j\in\mathcal{K}\backslash \{k\}}\delta_k\|\bw_{j,0}\|^2  \nonumber\\
+\; \bar{\mathbb{I}}_p + \sigma_k^2\Bigr)/\omega_k^{0}\leq  \vartheta_k^{10}, \; \forall k,\IEEEyessubnumber \label{eq:Rremain:c}\\
\Bigl(\sum_{j\in\mathcal{K}\backslash \{k\}}|\tilde{\bh}^{H}_{k}\bw_{j,1}|^2 + \sum_{j\in\mathcal{K}\backslash \{k\}}\delta_k\|\bw_{j,1}\|^2  \nonumber\\
+\; \bar{\mathbb{I}}_p+\sigma_k^2\Bigr)/\omega_k^{1}\leq  \vartheta_k^{11}, \; \forall k\IEEEyessubnumber \label{eq:Rremain:d}
\end{IEEEeqnarray}
where $\omega_k^{1}$, $\vartheta_k^{01}$, $\vartheta_k^{10}$, and $\vartheta_k^{11}$ are newly introduced variables.

From \eqref{eq:R00:6} and \eqref{eq:R01:concave}-\eqref{eq:R11:concave}, the objective \eqref{eq:eq:equiexpress:a} is transformed to the following concave
function:
\begin{IEEEeqnarray}{rCl}\label{eq:objconcave}
\mathcal{R}^{(n)}\bigl(\tau, \boldsymbol{\vartheta} \bigr) = \sum_{k=1}^K\mathcal{R}_k^{(n)}\bigl(\tau, \vartheta_k^{00}, \vartheta_k^{01}, \vartheta_k^{10}, \vartheta_k^{11} \bigr)
\end{IEEEeqnarray}
where $\boldsymbol{\vartheta}\triangleq\bigl[\vartheta_k^{00}, \vartheta_k^{01}, \vartheta_k^{10}, \vartheta_k^{11}\bigr]^T_{k\in\mathcal{K}}$, and
\begin{IEEEeqnarray}{rCl}\label{eq:Rateconcave}
\mathcal{R}_k^{(n)}\bigl(\tau, \vartheta_k^{00}, \vartheta_k^{01}, \vartheta_k^{10}, \vartheta_k^{11} \bigr) =  \widetilde{\mathcal{P}}_{00}\widetilde{\mathcal{R}}_{k}^{00,(n)}(\tau, \vartheta_k^{00} )  \nonumber\\
+\;  \widetilde{\mathcal{P}}_{01}\widetilde{\mathcal{R}}_{k}^{01,(n)}(\tau, \vartheta_k^{01} ) + \widetilde{\mathcal{P}}_{10}\widetilde{\mathcal{R}}_{k}^{10,(n)}(\tau, \vartheta_k^{10} )\nonumber\\
  +    \widetilde{\mathcal{P}}_{11} \widetilde{\mathcal{R}}_{k}^{11,(n)}(\tau, \vartheta_k^{11} ).
\end{IEEEeqnarray}

\textit{Convex approximation of the nonconvex constraint \eqref{eq:eq:equiexpress:b}:} We now turn our attention to \eqref{eq:eq:equiexpress:b}. As a result for \eqref{eq:Rateconcave},  the constraint \eqref{eq:eq:equiexpress:b} is inner approximated by the following linear  constraint:
\begin{IEEEeqnarray}{rCl}\label{eq:eq:equiexpress:1b}
\mathcal{R}_k^{(n)}\bigl(\tau, \vartheta_k^{00}, \vartheta_k^{01}, \vartheta_k^{10}, \vartheta_k^{11} \bigr) \geq \mathsf{\bar{R}}_{k},\;\forall k.
\end{IEEEeqnarray}
Thus, the convex program provides minorant maximization solved at the $(n + 1)$-th iteration  for the nonconvex program \eqref{eq:equiexpress:1}, as given by
\begin{IEEEeqnarray}{rCl}\label{eq:convexappro:1}
&&\underset{\bw_{0},\bw_{1},\tau,\boldsymbol{\omega},\boldsymbol{\vartheta}}{\mathrm{maximize}}\  \mathcal{R}^{(n)}\bigl(\tau, \boldsymbol{\vartheta} \bigr)       \IEEEyessubnumber \label{eq:eq:convexappro:a}\\
&&                          \st \,  \mathcal{R}_k^{(n)}\bigl(\tau, \vartheta_k^{00}, \vartheta_k^{01}, \vartheta_k^{10}, \vartheta_k^{11} \bigr) \geq \mathsf{\bar{R}}_{k},\forall k \in\mathcal{K},\IEEEyessubnumber \label{eq:convexappro:b} \qquad\\
											&&\qquad\qquad\ \eqref{eq:eq:equiexpress:c}, \eqref{eq:eq:equiexpress:d}, \eqref{eq:eq:equiexpress:e},\eqref{eq:trustregion},\eqref{eq:R00:7},\eqref{eq:trustregions},\eqref{eq:Rremain}\IEEEyessubnumber \label{eq:convexappro:e}
\end{IEEEeqnarray}
where $\boldsymbol{\omega}\triangleq\bigl[\omega_k^{0}, \omega_k^{1}\bigr]^T_{k\in\mathcal{K}}$. Note that the feasible set of \eqref{eq:convexappro:1} is also feasible for \eqref{eq:equiexpress:1}. We outline the proposed method in Algorithm~\ref{algo:proposed:BF}.
\begin{algorithm}[t]
\begin{algorithmic}[1]
\protect\caption{Proposed iterative algorithm to solve \eqref{eq:equiexpress:1}}
\label{algo:proposed:BF}
\global\long\def\algorithmicrequire{\textbf{Initialization:}}
\REQUIRE  Set $n:=0$ and solve \eqref{eq:initial:1} to generate an initial feasible point $\bigl(\bw_{0}^{(0)},\bw_{1}^{(0)},\tau^{(0)}\bigr)$.
\REPEAT
\STATE Solve \eqref{eq:convexappro:1} to obtain the optimal solution: $(\bw_{0}^{*},\bw_{1}^{*},\tau^{*},\boldsymbol{\omega}^{*},\boldsymbol{\vartheta}^{*})$.
\STATE Update\ $\bw_{0}^{(n+1)} := \bw_{0}^{*},\; \bw_{1}^{(n+1)} := \bw_{1}^{*},\; \tau^{(n+1)} := \tau^{*}$.
\STATE Set $n:=n+1.$
\UNTIL Convergence\\
\end{algorithmic} \end{algorithm}
After finding the optimal solution (step 2), we update the involved variables (step 3) and repeatedly solve \eqref{eq:convexappro:1} until convergence.

\textit{Generation of initial points:}
In fact, Algorithm~\ref{algo:proposed:BF} requires a feasible point of \eqref{eq:equiexpress:1} to meet the nonconvex throughput constraints, which is difficult to find in general. To overcome this issue, we successively solve  the following problem:
\begin{IEEEeqnarray}{rCl}\label{eq:initial:1}
\underset{\bw_{0},\bw_{1},\tau,\boldsymbol{\omega},\boldsymbol{\vartheta}}{\mathrm{maximize}}\ &&\underset{k\in\mathcal{K}}{\min}\ \left\{\mathcal{R}_k^{(n)}\bigl(\tau, \vartheta_k^{00}, \vartheta_k^{01}, \vartheta_k^{10}, \vartheta_k^{11} \bigr) - \mathsf{\bar{R}}_{k} \right\}      \IEEEyessubnumber \label{eq:initial:a}\qquad\  \\
                          \st \;&&  \eqref{eq:eq:equiexpress:c}, \eqref{eq:eq:equiexpress:d}, \eqref{eq:eq:equiexpress:e},\eqref{eq:trustregion},\eqref{eq:R00:7},\eqref{eq:trustregions},\eqref{eq:Rremain}\IEEEyessubnumber \label{eq:initial:b}
\end{IEEEeqnarray}
which is initialized by any feasible point $\bigl(\bw_{0}^{(0)},\bw_{1}^{(0)},\tau^{(0)}\bigr)$. We solve \eqref{eq:initial:1} until reaching
\begin{equation}
\underset{k\in\mathcal{K}}{\min}\ \left\{\mathcal{R}_k^{(n)}\bigl(\tau, \vartheta_k^{00}, \vartheta_k^{01}, \vartheta_k^{10}, \vartheta_k^{11} \bigr) - \mathsf{\bar{R}}_{k} \right\} \geq 0.\nonumber
\end{equation}

\begin{proposition}\label{prop1} Algorithm~\ref{algo:proposed:BF} returns a better point $\bigl(\bw_{0}^{(n)},\bw_{1}^{(n)},\tau^{(n)}\bigr)$ of \eqref{eq:equiexpress:1} and \eqref{eq:convexappro:1} after every iteration.  Hence, Algorithm~\ref{algo:proposed:BF} generates a non-decreasing sequence of  objective values and  also converges to a KKT  point of \eqref{eq:equiexpress:1} after a finite number of iterations.
\end{proposition}

\begin{IEEEproof}  Here we  provide a sketch of the proof to verify the statement. For ease of reference, let us define the objectives of \eqref{eq:equiexpress:1} and \eqref{eq:convexappro:1} w.r.t. the  updated optimization variables as $\mathcal{O}\bigl(\bw_{0},\bw_{1},\tau\bigr)$ and $\mathcal{O}^{(n)}\bigl(\bw_{0},\bw_{1},\tau\bigr)$, respectively. We know that $\mathcal{O}\bigl(\bw_{0},\bw_{1},\tau\bigr) \geq \mathcal{O}^{(n)}\bigl(\bw_{0},\bw_{1},\tau\bigr)$ (due to \eqref{eq:R00:2}) and $\mathcal{O}\bigl(\bw_{0}^{(n)},\bw_{1}^{(n)},\tau^{(n)}\bigr) = \mathcal{O}^{(n)}\bigl(\bw_{0}^{(n)},\bw_{1}^{(n)},\tau^{(n)}\bigr)$   (due to \eqref{eq:R00:5}). Therefore, we have
\begin{equation}\begin{aligned}\nonumber
&\mathcal{O}\bigl(\bw_{0}^{(n+1)},\bw_{1}^{(n+1)},\tau^{(n+1)}\bigr) \geq \\
&\mathcal{O}^{(n)}\bigl(\bw_{0}^{(n+1)},\bw_{1}^{(n+1)},\tau^{(n+1)}\bigr) \geq\\
& \mathcal{O}^{(n)}\bigl(\bw_{0}^{(n)},\bw_{1}^{(n)},\tau^{(n)}\bigr) = \mathcal{O}\bigl(\bw_{0}^{(n)},\bw_{1}^{(n)},\tau^{(n)}\bigr).
\end{aligned}\end{equation}
This implies that $\bigl(\bw_{0}^{(n+1)},\bw_{1}^{(n+1)},\tau^{(n+1)}\bigr)$ is a better point for \eqref{eq:equiexpress:1} than $\bigl(\bw_{0}^{(n)},\bw_{1}^{(n)},\tau^{(n)}\bigr)$. Hence, $\{\mathcal{O}\bigl(\bw_{0}^{(n)},\bw_{1}^{(n)},$ $\tau^{(n)}\bigr)\}_{n\geq 1}$ is a non-decreasing sequence and possibly converges to positive infinity. However, this sequence is bounded above due to the power constraint \eqref{eq:eq:equiexpress:c}. As $n$ tends to infinity, Algorithm~\ref{algo:proposed:BF} converges to an accumulation point $\bigl(\bar{\bw}_{0},\bar{\bw}_{1},\bar{\tau}\bigr)$, i.e., $\underset{n\rightarrow +\infty}{\lim}\mathcal{O}\bigl(\bw_{0}^{(n)},\bw_{1}^{(n)},$ $\tau^{(n)}\bigr) = \mathcal{O}\bigl(\bar{\bw}_{0},\bar{\bw}_{1},\bar{\tau}\bigr)$. Thus, we can prove that Algorithm~\ref{algo:proposed:BF} converges to a KKT point of \eqref{eq:equiexpress:1} according to \cite[Theorem 1]{MW78}. Furthermore,  Algorithm~\ref{algo:proposed:BF} will terminate after a finite number of iterations when it satisfies
\[\left| \frac{\mathcal{O}\bigl(\bw_{0}^{(n+1)},\bw_{1}^{(n+1)},\tau^{(n+1)}\bigr)-\mathcal{O}\bigl(\bw_{0}^{(n)},\bw_{1}^{(n)},\tau^{(n)}\bigr)}{\mathcal{O}\bigl(\bw_{0}^{(n)},\bw_{1}^{(n)},\tau^{(n)}\bigr)}\right| \leq \epsilon_{\mathtt{err}}
\]
where $\epsilon_{\mathtt{err}} > 0$ is a given tolerance.  Proposition \ref{prop1} is thus proved.
\end{IEEEproof}

\textit{Complexity Analysis}:  The computational complexity
of solving convex problem \eqref{eq:convexappro:1}  is owing to only   simple convex quadratic and linear constraints at each iteration of Algorithm~\ref{algo:proposed:BF}.   To be specific,  the convex problem \eqref{eq:convexappro:1} has  $ (2N_t+6)K + 1$ real-valued scalar decision variables, a linear objective,  $3K+1$ linear constraints,  and $8K+M+1$ quadratic constraints. Then,  the computational complexity per iteration to solve \eqref{eq:convexappro:1} is $\mathcal{O}\Bigl(\bigl((2N_t+6)K + 1\bigr)^2\sqrt{11K+M+2}(2N_tK+17K+M+3) \Bigr)$ \cite{Nesterov94}.

\section{Numerical Results}
\label{Numericalresults}

We now evaluate the performance of the proposed design using computer simulations. The entries of $\mathbf{h}_k,\forall k\in\mathcal{K}$ and $\mathbf{g}_m,\forall m\in\mathcal{M}$ are assumed to undergo the effects of large-scale path loss and small-scale fading. Specifically, we set the path loss exponent as $\mathtt{PL} = 3$. Small-scale fading is then generated as Rician fading with the Rician factor  $K_{\mathtt{R}} = 10$ dB \cite{ShiTWC14}. The maximum interference power constraints  at all PUs and  minimum rate constraints for all SUs are set to be equal, i.e., $\mathcal{I}_m = \mathcal{I}$, $\forall m\in\mathcal{M}$ and $\mathsf{\bar{R}}_k = \mathsf{\bar{R}}$, $\forall k\in\mathcal{K}$. For the results of a local prediction, the probabilities for a wrong
prediction and  successful prediction for  the true state of PUs are set to $\mathcal{P}_p^w = 0.25$ and $\mathcal{P}_p^s = 0.7$, respectively \cite{TumuluruWCMC12}. For spectrum sensing, the target detection probability is set to $\mathcal{P}_d=\mathcal{\bar{P}}_d = 0.9$, which meets  the requirements for IEEE 802.22 with {\color{black}a low SNR of} $\gamma = -15$ dB and $f_s = 1500$ samples/s \cite{IEEE80222}.
Unless stated otherwise, the other parameters  given in Table~\ref{parameter}   follow those obtained from \cite{LiangTWC08, TumuluruWCMC12,CichonCST16,KangTVT09}. In Table~\ref{parameter}, we  assume that the secondary BS can achieve better channel estimates for their serving  users (SUs) compared to the PUs. {\color{black}The error tolerance between two consecutive iterations in Algorithm 1 is set to $\epsilon_{\mathtt{err}}  = 10^{-3}$}.
We divide the achieved  sum rate  by $\ln(2)$ to arrive at a unit of bps/channel-use  over an average of 10,000 simulated slots.

\begin{table}[t]
\caption{Simulation Parameters}
	\label{parameter}
	\centering
	{\setlength{\tabcolsep}{0.6em}
\setlength{\extrarowheight}{0.3em}
	\begin{tabular}{l|l}
		\hline
				Parameter & Value \\
		\hline\hline
		    Radius of considered cell, $r$                    & 100 m \\
				Distance between the SBS and the nearest user  & $\geq$ 10 m\\
				Noise variances, $\sigma^2_{k}$& -90 dBm \\
				Predetermined rate threshold, $\bar{\mathsf{R}}$ & 0.5 bps/Hz\\
				Predetermined interference power constraint, $\mathcal{I}$ & -5 dBm\\
				Averaged interference at the SUs, $\bar{\mathbb{I}}_p$ & 5 dBm \\
				Number of antennas at the SBS, $N_t$ & 8\\
				Normalized uncertainties of SUs' channel, $\epsilon_s$& $10^{-3}$\\
				Normalized uncertainties of PUs' channel, $\epsilon_p$& $10^{-2}$\\
				Slot length, $T$ & 100 ms \\
				Prediction time at SUs and SBS, $t_p$& 5 ms\\
				Time for $\mathsf{SU}_k$ to  report to FC, $t_{r_k}\,\forall k$& 0.2 ms\\
				Time for FC to make a decision, $t_{\mathtt{FC}}$ & 1 ms\\
		\hline		   				
		\end{tabular}}
\end{table}

\begin{figure}
    \begin{center}
    \begin{subfigure}[Probability of miss-detection versus traffic intensity.]{
        \includegraphics[width=0.43\textwidth,trim={0.0cm -0.0cm 0.0cm -0.0cm}]{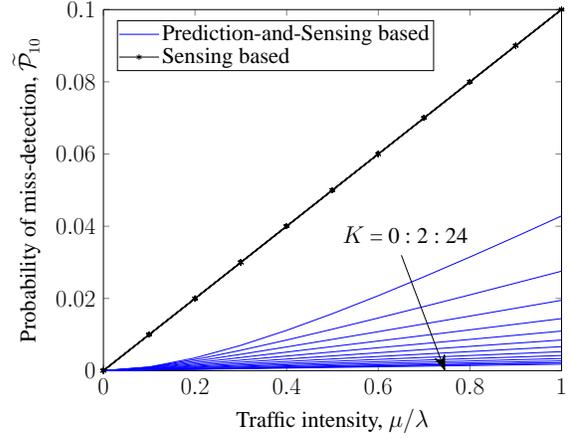}}
    		\label{fig:Protran:a}
				\end{subfigure}
		\begin{subfigure}[Probability of  detection versus traffic intensity.]{
        \includegraphics[width=0.42\textwidth,trim={0.0cm -0.0cm 0.0cm -0.0cm}]{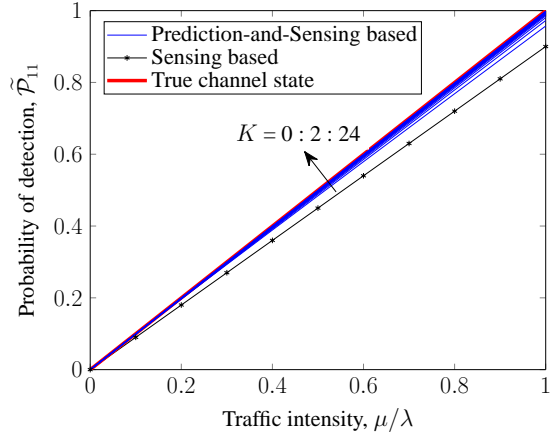}}
        \label{fig:fig:Protran:b}
    \end{subfigure}
		\caption{Probabilities of (a) miss-detection  and (b) detection  versus traffic intensity.}\label{fig:Protran}
\end{center}
\end{figure}

We evaluate the  probabilities of miss-detection $\mathcal{\widetilde{P}}_{10}$ in Fig.~\ref{fig:Protran}(a) and detection $\mathcal{\widetilde{P}}_{11}$ in Fig.~\ref{fig:Protran}(b) versus the traffic intensity. We also compare  the corresponding probabilities to   spectrum sensing only \cite{LiangTWC08}. For  spectrum sensing only, the probabilities of miss-detection and detection are calculated as $\mathcal{\widetilde{P}}_{10} = \Pro(\mathcal{H}_1)(1-\mathcal{P}_d)$ and $\mathcal{\widetilde{P}}_{11} = \Pro(\mathcal{H}_1)\mathcal{P}_d$, respectively. The proposed prediction-and-sensing scheme achieves better  performance than sensing only in all cases, and its gain is even deeper when the intensity of the traffic  increases. In addition, increasing  the number of SUs leads to a reduction in $\mathcal{\widetilde{P}}_{10}$ and increase in $\mathcal{\widetilde{P}}_{11}$. Specifically, the  probability of miss-detection $\mathcal{\widetilde{P}}_{10}$ is always less than 5$\%$, and the probability of  detection $\mathcal{\widetilde{P}}_{11}$ is very close to the value corresponding to the true channel state when $K$ = 24.

\begin{figure}[t]
\centering
        \includegraphics[width=0.45\textwidth]{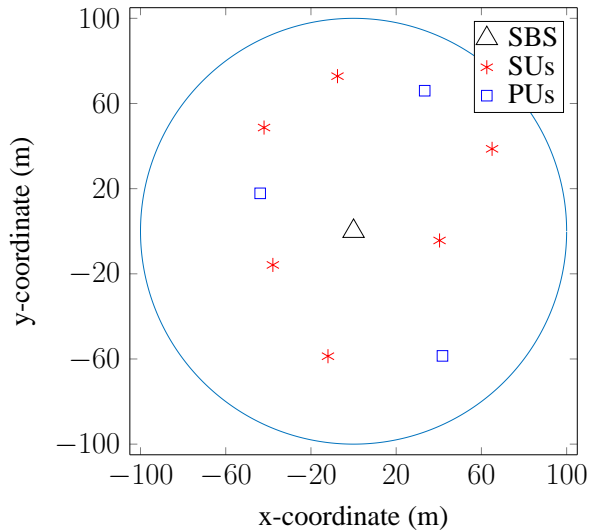}
	  \caption{Location of users of the simulation setup with $K$ = 6 and $M$ = 3.}\label{fig:celllayout}
\end{figure}

We will use the simulation setup illustrated in Fig.~\ref{fig:celllayout} to evaluate the system performance in terms of the sum rate. We also compare the  sum rate of the proposed PSBSS to that of \textit{spectrum underlay} \cite{Zhang08,ZhangTSP08,NguyenTVT16,NguyenCL16,HeTWC14,Zhang09} and \textit{opportunistic spectrum access} \cite{GanTVT12,PehTVT09,ZhaoSPM07,ZhaoWC07,LiangTWC08}. In particular, we consider the following optimization problems:
\begin{IEEEeqnarray}{rCl}\label{eq:opportunistic:1}
\underset{\bw_{0},\, t_s}{\mathrm{maximize}}\ &&  \Bigl(1 - \frac{t_{pr}+t_s}{T}\Bigr)\sum\nolimits_{k=1}^K\Bigl[ \widetilde{\mathcal{P}}_{00}R_{k}^{00}(\bw_{0}) \nonumber\\
&&\qquad\qquad\qquad\qquad +\; \widetilde{\mathcal{P}}_{10}R_{k}^{10}(\bw_{0})\Bigr]
 \IEEEyessubnumber \label{eq:opportunistic:a}\\
                          \st \;&& \Bigl(1 - \frac{t_{pr}+t_s}{T}\Bigr) \Bigl[ \widetilde{\mathcal{P}}_{00}R_{k}^{00}(\bw_{0}) \nonumber\\
													&&\qquad\quad+\; \widetilde{\mathcal{P}}_{10}R_{k}^{10}(\bw_{0})\Bigr]
  \geq \mathsf{\bar{R}}_{k},\;\forall k \in\mathcal{K},\IEEEyessubnumber \label{eq:opportunistic:b} \\
							&&\Bigl(1 - \frac{t_{pr}+t_s}{T}\Bigl)\sum\nolimits_{k=1}^K \widehat{\mathcal{P}}_0\|\bw_{k,0}\|^2 
									\leq P_{sbs},\IEEEyessubnumber \label{eq:opportunistic:d}\qquad\quad\\
							&& \eqref{eq:problem:c}\IEEEyessubnumber
\end{IEEEeqnarray}
for the opportunistic spectrum access model  and
\begin{IEEEeqnarray}{rCl}\label{eq:underlay:1}
&&\underset{\bw_{1}}{\mathrm{maximize}}\  \sum\nolimits_{k=1}^KR_{k}^{11}(\bw_{1})  \IEEEyessubnumber \label{eq:underlay:a}\\
                         && \st \; R_{k}^{11}(\bw_{1})  \geq \mathsf{\bar{R}}_{k},\;\forall k \in\mathcal{K},\IEEEyessubnumber \label{eq:underlay:b} \\
							&&\sum\nolimits_{k=1}^K \|\bw_{k,1}\|^2\leq P_{sbs},\IEEEyessubnumber \label{eq:underlay:d}\\
							&&  \sum_{k=1}^K\Bigl(\|\tilde{\mathbf{g}}^H_m\bw_{k,1}\|^2 + \hat{\delta}_m\|\bw_{k,1}\|^2\Bigr)\leq \mathcal{I}_m,\forall m\in\mathcal{M}\IEEEyessubnumber \label{eq:underlay:e} \quad
\end{IEEEeqnarray}
for the spectrum underlay model. It is obvious that the problems \eqref{eq:opportunistic:1} and \eqref{eq:underlay:1} can also be solved using Algorithm~\ref{algo:proposed:BF}. In what follows, the probability when the PUs' channel is idle $\Pro(\mathcal{H}_0)$ is set to $\Pro(\mathcal{H}_0) = 0.6$, following the guidelines provided by the FCC \cite{FCC},  unless specified otherwise.

\begin{figure}[t]
\centering
        \includegraphics[width=0.45\textwidth]{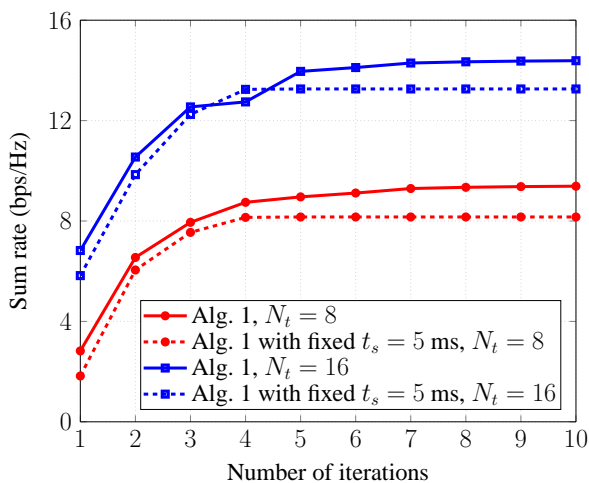}
	  \caption{Convergence behavior  of Algorithm~\ref{algo:proposed:BF} with $\Pro(\mathcal{H}_0) = 0.6$ and $P_{sbs} = 20$ dBm.}\label{fig:Convergencebehavior}
\end{figure}

In Fig.~\ref{fig:Convergencebehavior}, we illustrate the convergence behavior of Algorithm~\ref{algo:proposed:BF} with different numbers of transmit antennas, $N_t\in\{8, 16\}$ for one random channel realization. We can see that Algorithm~\ref{algo:proposed:BF} with joint optimization converges very fast to reach its optimal solution. Specifically,  it  converges within 8 iterations and is insensitive to an increase in $N_t$. We also observe that if
 the sensing time is fixed to $t_s = 5$ ms, Algorithm~\ref{algo:proposed:BF}  converges more quickly in about 5 iterations, but the corresponding sum rates are worse than with joint optimization. The slower convergence for joint optimization in Algorithm~\ref{algo:proposed:BF}  is probably   due to coupling between the beamforming vectors and $t_s$ in both the objective and the constraints.
As expected,  we obtain a higher sum rate with a larger number of transmit antennas. {\color{black}On the average, Algorithm \ref{algo:proposed:BF} requires about
8.3 iterations with $N_t = 8$ and 8.7 iterations with $N_t = 16$ for convergence.}

\begin{figure}[t]
\centering
        \includegraphics[width=0.45\textwidth]{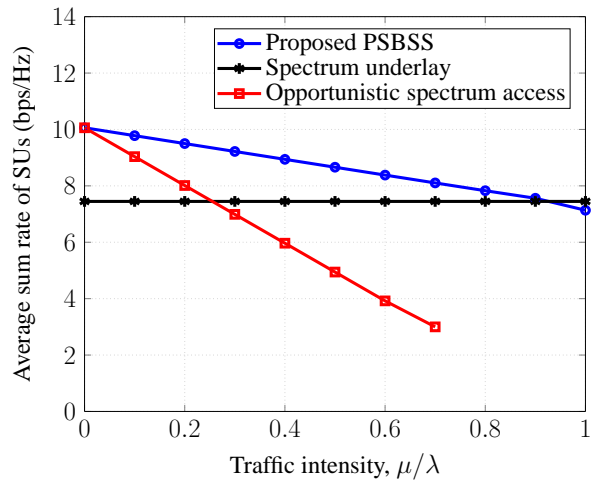}
	  \caption{Average sum rate of SUs versus traffic intensity with $P_{sbs} = 20$ dBm. }\label{fig:SRvsTraffic}
\end{figure}

Fig.~\ref{fig:SRvsTraffic} depicts the effect of the traffic intensity, $\mu/\lambda$, on the system performance. An increase in traffic intensity leads to a decrease in the sum rate of the opportunistic spectrum access model since the opportunity for SUs to access spectrum resources is accordingly reduced. In contrast, the sum rate of the spectrum underlay model is independent of the traffic intensity, which can be easily verified from \eqref{eq:underlay:1}. When $\mu/\lambda < 0.9$, the proposed PSBSS model outperforms the others in terms of the sum rate because it exploits the advantages of both models. Note that for $\mu/\lambda \geq 0.9$, the sum rate of the proposed PSBSS model tends to be worse than the spectrum underlay one. In this case, the proposed PSBSS actually becomes the spectrum underlay, but it still needs to expend time resources  to detect the channels.

\begin{figure}
    \begin{center}
    \begin{subfigure}[Average sum rate of SUs versus $P_{sbs}$.]{
        \includegraphics[width=0.45\textwidth,trim={0.0cm -0.0cm 0.0cm -0.00cm}]{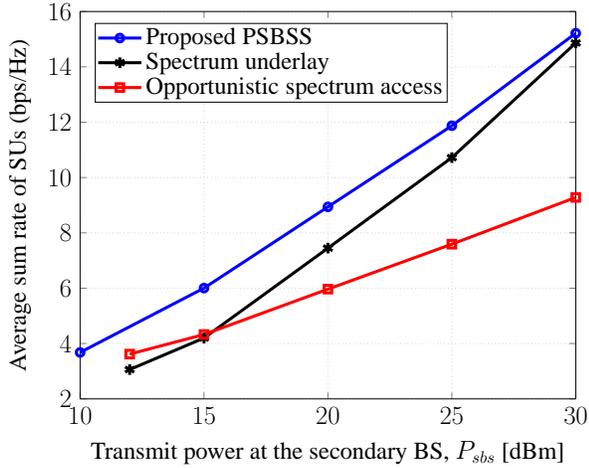}}
    		\label{fig:SRvsPsbs}
				\end{subfigure}
		\begin{subfigure}[Average sum rate of SUs versus $\mathcal{I}$ with $P_{sbs} = 20$ dBm.]{
        \includegraphics[width=0.45\textwidth,trim={0.0cm -0.0cm 0.0cm -0.00cm}]{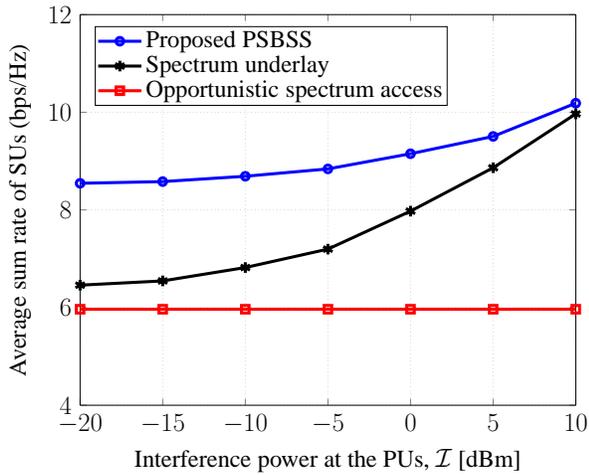}}
        \label{fig:SRvsIatPUs}
    \end{subfigure}
		\caption{Average sum rate of SUs (a) versus the transmit power constraint at the secondary BS and (b) versus the interference power constraint at the PUs with $\Pro(\mathcal{H}_0) = 0.6$.}\label{fig:SRvsPsbsI}
\end{center}
\end{figure}

We plot the sum rate of the secondary system versus the transmit power constraint, $P_{sbs}$, in Fig.~\ref{fig:SRvsPsbsI}(a) and the interference power constraint, $\mathcal{I}$, in Fig.~\ref{fig:SRvsPsbsI}(b). In general, the sum rates of all models increase with higher $P_{sbs}$ and $\mathcal{I}$, except the opportunistic spectrum access in Fig.~\ref{fig:SRvsPsbsI}(b) because this model transmits without the effect of the interference power constraint and is limited only by the transmit power given in \eqref{eq:opportunistic:1}. The sum rate of the spectrum underlay  can be very close to that of the proposed PSBSS at the peaks $P_{sbs} = 30$ dBm and $\mathcal{I} = 10$ dBm. As shown in both Fig.~\ref{fig:SRvsPsbsI}(a) and Fig.~\ref{fig:SRvsPsbsI}(b), the transmit power has greater influence on the sum rates of all models than the interference power. Moreover, the sum rates of the proposed PSBSS are always larger than those of the others, which further confirms the superiority of the proposed method.

\begin{figure}[t]
\centering
        \includegraphics[width=0.45\textwidth]{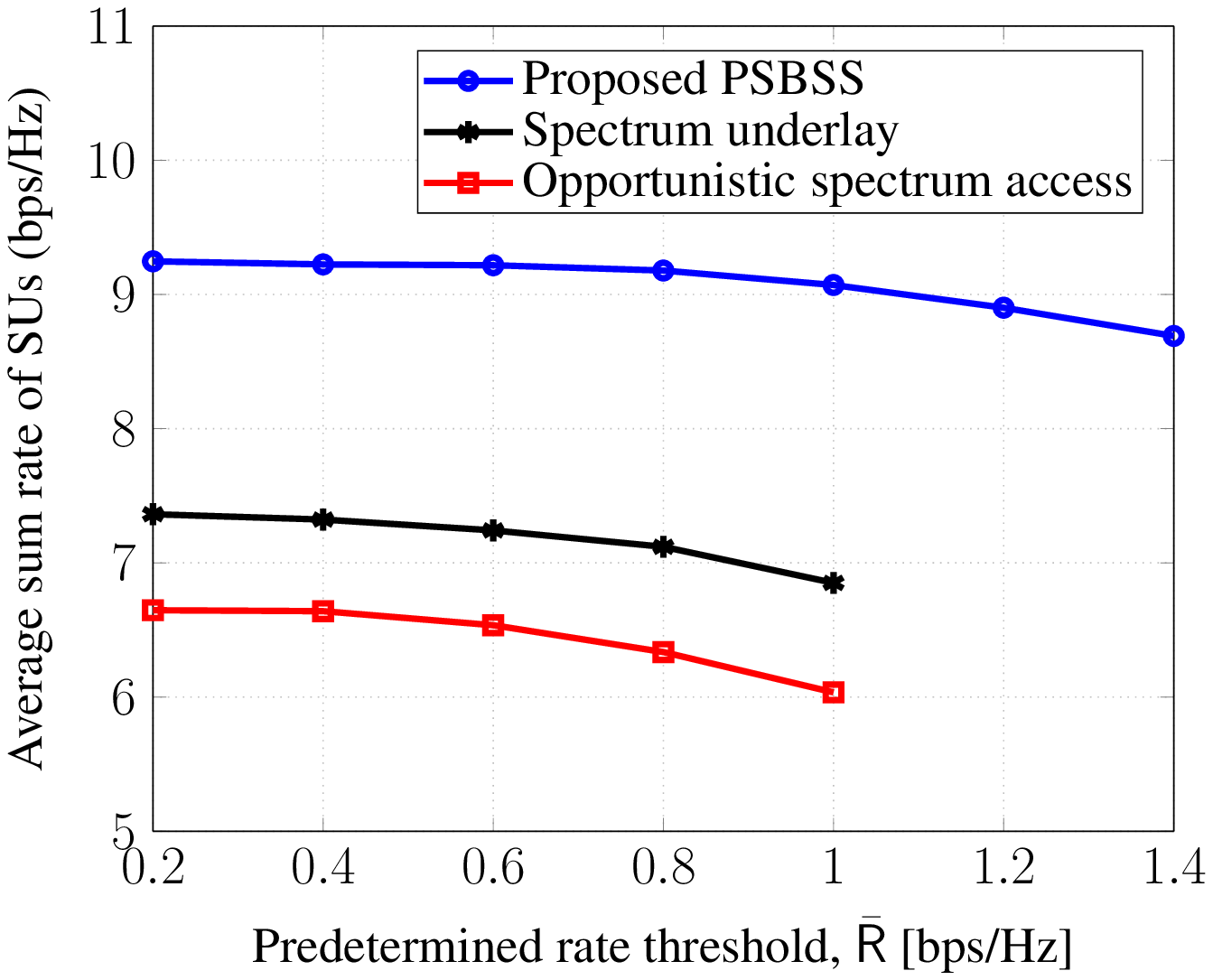}
	  \caption{Average sum rate of SUs versus $\bar{\mathsf{R}}$ with $\Pro(\mathcal{H}_0) = 0.6$ and $P_{sbs} = 20$ dBm. }\label{fig:SRvsR}
\end{figure}

The sum rate  versus the predetermined rate threshold $\bar{\mathsf{R}}$ bps/Hz is shown in Fig.~\ref{fig:SRvsR}. Certainly, the sum rates of all models  monotonically decrease for $\bar{\mathsf{R}}\in[0.2,\;1.4]$ bps/Hz due to  the secondary BS  paying more attention to serving SUs
with poor channel conditions by transferring more power to them  and scaling down the power transmitted to SUs with good channel conditions. Again, the proposed PSBSS outperforms the others in terms of the sum rate for all ranges of $\bar{\mathsf{R}}$. Another interesting observation is that
the spectrum underlay and opportunistic spectrum access are infeasible  for $\bar{\mathsf{R}} > 1$ bps/Hz, i.e.,
these models cannot offer such high rate threshold to all SUs. In contrast, the proposed model is still feasible at $\bar{\mathsf{R}} = 1.4$ bps/Hz and achieves less degradation than the others, which indicates the robustness of our proposed Algorithm~\ref{algo:proposed:BF}.

\begin{figure}[t]
\centering
        \includegraphics[width=0.45\textwidth]{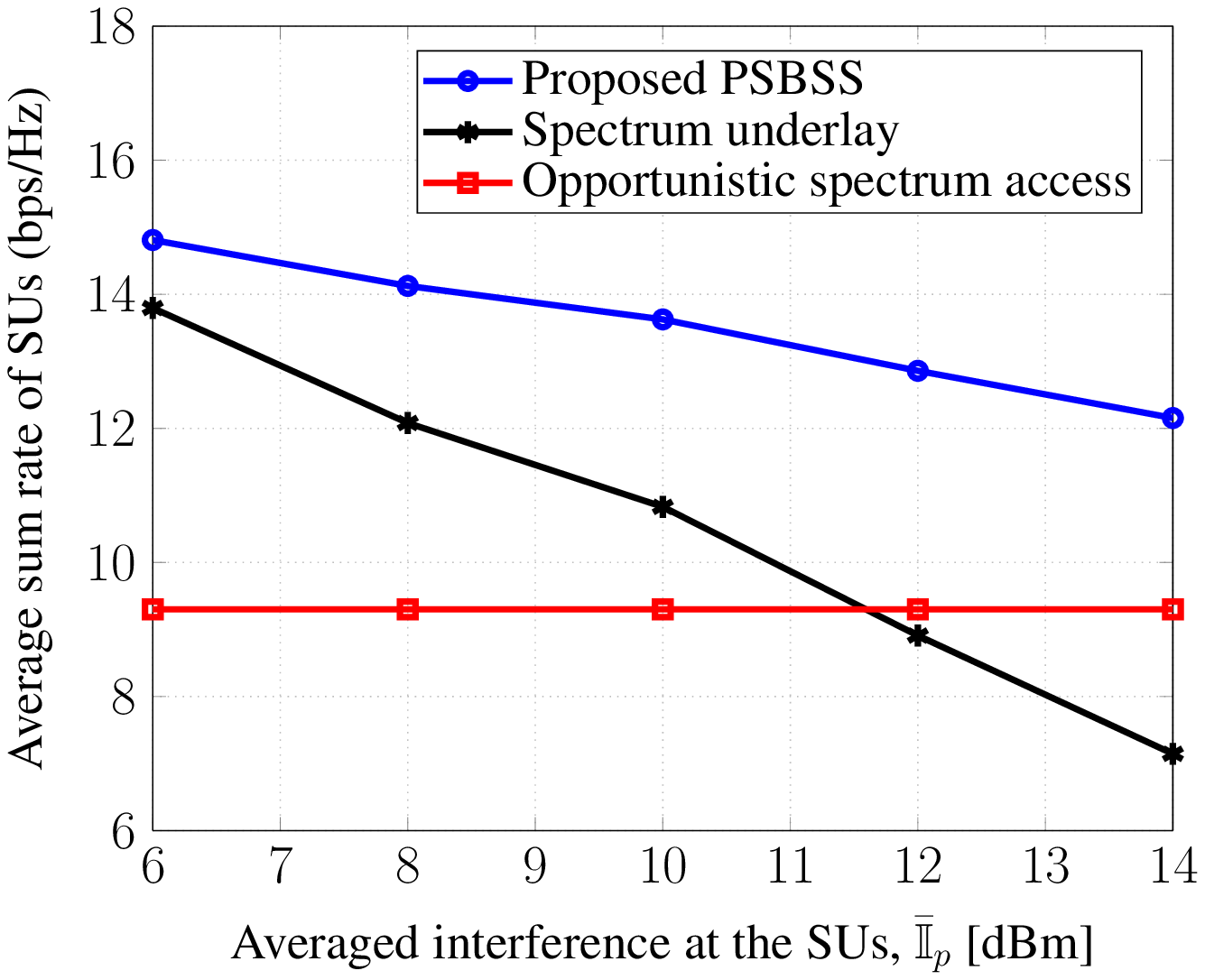}
	  \caption{Average sum rate of SUs versus $\bar{\mathbb{I}}_p$ with $\Pro(\mathcal{H}_0) = 0.6$ and $P_{sbs} = 30$ dBm. }\label{fig:SRvsIp}
\end{figure}

In Fig.~\ref{fig:SRvsIp}, we examine the effect of the interference $\bar{\mathbb{I}}_p$ caused by the primary system. As can be seen, an increase in  $\bar{\mathbb{I}}_p$ results  in a dramatic degradation of the sum rate of  the spectrum underlay. We should emphasize that though   the opportunistic spectrum access model depends on $\bar{\mathbb{I}}_p$ (i.e., $R_{k}^{10}(\bw_{0})$ given in \eqref{eq:opportunistic:1}), the resulting sum rate is nearly unchanged, even for a high $\bar{\mathbb{I}}_p$, because the probability of transmission for $R_{k}^{10}(\bw_{0})$ is negligible as $\widetilde{\mathcal{P}}_{10} \approx 0.6\%$. For whatever level of  $\bar{\mathbb{I}}_p$, the proposed PSBSS still  achieves a better sum rate than the others.

\begin{figure}
    \begin{center}
    \begin{subfigure}[Average sum rate of SUs versus the normalized uncertainty level associated to the PUs with $\epsilon_s = 10^{-3}$.]{
        \includegraphics[width=0.45\textwidth,trim={0.0cm -0.0cm 0.0cm -0.0cm}]{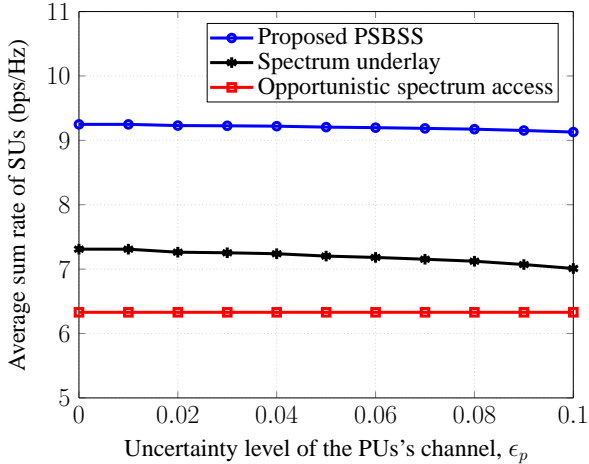}}
    		\label{fig:SRvsEpsilonPU:a}
				\end{subfigure}
		\begin{subfigure}[Average sum rate of SUs versus the normalized uncertainty level associated to the SUs with $\epsilon_p = 10^{-2}$.]{
        \includegraphics[width=0.45\textwidth,trim={0.0cm -0.0cm 0.0cm -0.0cm}]{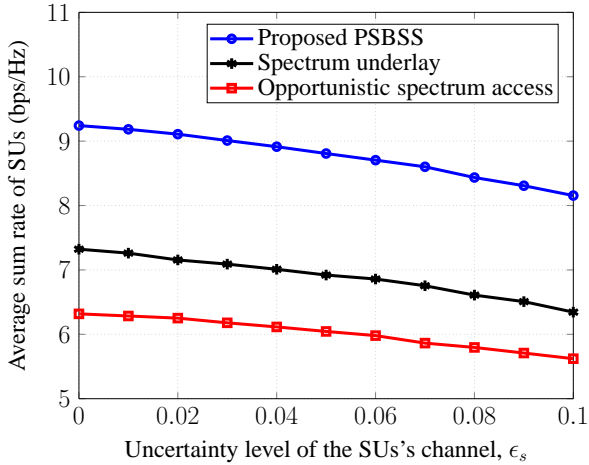}}
        \label{fig:SRvsEpsilonSU:b}
    \end{subfigure}
		\caption{Average sum rate of SUs (a) versus the normalized uncertainty level associated to the PUs and (b) versus the normalized uncertainty level associated to the SUs with $\Pro(\mathcal{H}_0) = 0.6$ and $P_{sbs} = 20$ dBm.}\label{fig:SRvsEpsilonSUPU}
\end{center}
\end{figure}

Next, we plot the sum rate against the channel uncertainty  of the PUs in Fig.~\ref{fig:SRvsEpsilonSUPU}(a) and the channel uncertainty  of the SUs in Fig.~\ref{fig:SRvsEpsilonSUPU}(b). In Fig.~\ref{fig:SRvsEpsilonSUPU}(a), the secondary system for fixed $\epsilon_s = 10^{-3}$ for all models has a very minimal loss on the  sum rates achieved when $\epsilon_p$ increases. In Fig.~\ref{fig:SRvsEpsilonSUPU}(b), the channel uncertainty  of SUs $\epsilon_s$ for fixed $\epsilon_p = 10^{-2}$ has a visible effect on the achieved sum rates, especially for a higher $\epsilon_s$. Herein,  an  important engineering insight is that the sum rate of the secondary system is more sensitive to the estimation errors for the SUs' channels than for those of  the PUs' channels.

\begin{figure}[t]
\centering
        \includegraphics[width=0.45\textwidth]{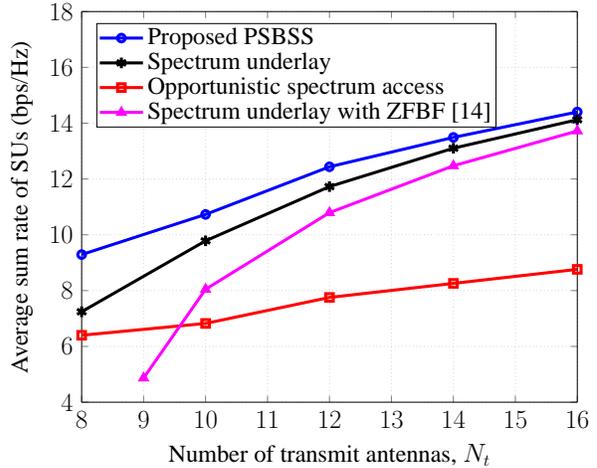}
	  \caption{Average sum rate of SUs versus the number of transmit antennas at the secondary BS for perfect channel estimation ($\epsilon_s=\epsilon_p=0$) with $\Pro(\mathcal{H}_0) = 0.6$ and $P_{sbs} = 20$ dBm. }\label{fig:SRvsNt}
\end{figure}

Finally, Fig.~\ref{fig:SRvsNt} compares the sum  rate performance of the proposed PSBSS to that of the spectrum underlay, opportunistic spectrum access, and spectrum underlay with zero-forcing beammforming (ZFBF) \cite{NguyenTVT16}. For the spectrum underlay with ZFBF in \cite{NguyenTVT16}, the secondary BS places null spaces at the beamforming vector of each  SU to cancel co-channel interference. Since a perfect CSI has been assumed  in \cite{NguyenTVT16}, thus to ensure a fair comparison between those models,  we solve the proposed Algorithm~\ref{algo:proposed:BF} by assuming no channel uncertainty (i.e., $\epsilon_s = \epsilon_p = 0$). In Fig.~\ref{fig:SRvsNt}, we plot the sum rate  versus the number of antennas $N_t\in\{8,\;16\}$ for fixed $\Pro(\mathcal{H}_0) = 0.6$ and $P_{sbs} = 20$ dBm. For $N_t = 12$, we can clearly observe the gains about 0.71 bps/Hz, 1.64 bps/Hz, and 4.68 bps/Hz in the achieved sum  rate of the proposed PSBSS compared to that of the spectrum underlay, spectrum underlay with ZFBF \cite{NguyenTVT16}, and opportunistic spectrum access, respectively. In addition, the spectrum underlay with ZFBF is infeasible when $N_t < 9$ due to a lack of degrees of freedom to leverage multiuser diversity.  However, it yields a good sum rate performance for a large number of transmit antennas.

\section{Conclusions}
\label{Conclusion}
In this paper, we  proposed a prediction-and-sensing-based spectrum sharing model for cognitive radio networks. In this model,  the time structure of each resource block was redesigned to incorporate both spectrum prediction and spectrum sensing phases. Specifically,   simple cooperative spectrum prediction between all SUs and the secondary BS was proposed to help reduce the detection errors as well as improve the detection accuracy. We  studied the sum rate maximization problem considering the minimum rate requirements for each SU in the case where linear beamforming is adopted.
 To solve the original nonconvex optimization problem, we first transformed it into a more tractable form and then proposed a
 new iterative algorithm to maximize the sum rate of the secondary system.
The proposed design  captured all important factors in cognitive radio networks using a low-complexity algorithm.
The proposed algorithm with realistic parameters was numerically shown to have  fast convergence  almost independently of the problem size. 
 The sum rate of the proposed model was  thus  shown to be remarkably larger than that of conventional models. We also  discussed the effect of  the channel uncertainties for the SUs'  and PUs' channels. We concluded that the estimation error of the SUs' channels has a larger effect on the achievable sum rate of the secondary system than that of the PUs' channels.

\bibliographystyle{IEEEtran}
\balance
\bibliography{Journal}

\end{document}